\documentstyle[11pt,aaspp4]{article}

\begin{document} 

\hyphenation{PSPC}
\hyphenation{X-ray}

\slugcomment{To Appear in MNRAS}

\title{Sensitivity of Galaxy Cluster Morphologies to $\Omega_0$ and
$P(k)$}

\author{David A. Buote\altaffilmark{1}}
\affil{Department of Physics and Center for Space Research 37-241}
\affil{Massachusetts Institute of Technology}
\affil{77 Massachusetts Avenue, Cambridge, MA 02139, USA}
\affil{buote@ast.cam.ac.uk}

\author{Guohong Xu} 
\affil{Board of Studies in Astronomy and
Astrophysics,}
\affil{University of California at Santa Cruz,} 
\affil{Santa Cruz, CA 95064, USA}
\affil{xu@ucolick.org}

\altaffiltext{1}{Present Address: Institute of Astronomy, Madingley
Road, Cambridge CB3 0HA, UK}

\begin{abstract}
We examine the sensitivity of the spatial morphologies of galaxy
clusters to $\Omega_0$ and $P(k)$ using high-resolution N-body
simulations with large dynamic range. Variants of the standard CDM
model are considered having different spatial curvatures, SCDM
$(\Omega_0=1)$, OCDM $(\Omega_0=0.35)$, LCDM $(\Omega_0=0.35,
\lambda_0=0.65)$, and different normalizations, $\sigma_8$. We also
explore critical density models with different spectral indices, $n$,
of the scale-free power spectrum, $P(k)\propto k^n$.  Cluster X-ray
morphologies are quantified with power ratios (PRs), where we take for
the X-ray emissivity $j_{gas}\propto \rho^2_{DM}$, which we argue is a
suitable approximation for analysis of PRs.  We find that $\Omega_0$
primarily influences the means of the PR distributions whereas the
power spectrum ($n$ and $\sigma_8$) primarily affects their variances:
$\log_{10}(P_3/P_0)$ is the cleanest probe of $\Omega_0$ since its
mean is very sensitive to $\Omega_0$ but very insensitive to
$P(k)$. The PR means easily distinguish the SCDM and OCDM models,
while the SCDM and LCDM means show a more modest, but significant,
difference $(\sim 3\sigma)$.  The OCDM and LCDM models are largely
indistinguishable in terms of the PRs.  Finally, we compared these
models to a sample of $ROSAT$ clusters and find that the PR means of
the SCDM clusters exceed the $ROSAT$ means with a high formal level of
significance $(\sim 4\sigma)$.  Though the formal significance level
of this $\rho^2_{DM}$ / X-ray comparison should be considered only
approximate, we argue that taking into account the hydrodynamics and
cooling will not reconcile a discrepancy this large.  The PR means of
the OCDM clusters are consistent, and the means of the LCDM clusters
are marginally consistent, with those of the $ROSAT$ clusters.  Thus,
we conclude that cluster morphologies strongly disfavor $\Omega_0=1$,
CDM while favoring low density, CDM models with or without a
cosmological constant.
\end{abstract}


\section{Introduction} 

The quest for $\Omega_0$, the current ratio of the mean mass density
of the universe to the critical density required for closure, has been
a focus of the research efforts of many astrophysicists involving a
variety of different techniques. At present, most observational
evidence suggests a universe with sub-critical matter density, perhaps
with a cosmological constant making up the difference required for a
critical universe (e.g., Coles \& Ellis 1994; Ostriker \& Steinhardt
1995). The possibility of measuring $\Omega_0$ using the amount of
``substructure'' in galaxy clusters has thus generated some interest,
``This is a critical area for further research, as it directly tests
for $\Omega$ in dense lumps, so both observational and theoretical
studies on a careful quantitative level would be well rewarded.''
(Ostriker 1993).

Early analytical work (e.g., Richstone, Loeb, \& Turner 1992) and
simulations (Evrard et al. 1993; Mohr et al. 1995) found that the
morphologies of X-ray clusters strongly favored $\Omega_0\sim 1$ over
low-density universes. Along with $POTENT$ analysis of cosmic velocity
fields (e.g., Dekel 1994), these substructure analyses were the only
indicators in support of a critical value of $\Omega_0$.  However, the
analytical results (e.g., Kauffmann \& White 1993; Nakamura, Hattori,
\& Mineshige 1995), simulations (e.g., Jing et al. 1994), and
morphological statistics (e.g., Buote \& Tsai 1995b) have been
criticized rendering the previous conclusions about $\Omega_0$
uncertain.

Buote \& Tsai (1995b, hereafter BTa) introduced the power ratios (PRs)
for quantifying the spatial morphologies of clusters in terms of their
dynamical states. The PRs essentially measure the square of the ratio
of a higher order moment of the two-dimensional gravitational
potential to the monopole term computed within a circular aperture,
where the radius is specified by a metric scale (e.g., 1 Mpc). Buote
\& Tsai (1996, hereafter BTb) computed PRs of $ROSAT$ X-ray images for
a sample of 59 clusters and discovered that the clusters are strongly
correlated in PR space, obeying an ``evolutionary track'' which
describes the dynamical evolution of the clusters (in projection).
Tsai \& Buote (1996, hereafter TB) studied the PRs of a small sample
of clusters formed in the hydrodynamical simulation of Navarro, Frenk,
\& White (1995) and verified the interpretation of the ``evolutionary
track''. In contrast to the previous studies (e.g., Richstone et
al. 1992; Mohr et al. 1995), TB concluded that their small cluster
sample, formed in a standard $\Omega_0=1$, CDM simulation, possessed
too much substructure (as quantified by the PRs) with respect to the
$ROSAT$ clusters, and thus favored a lower value of $\Omega_0$.

However, a statistically large sample of clusters is important for
studies of cluster morphologies.  The PRs are most effective at
categorizing clusters into different broad morphological types;
i.e. the distinction between equal-sized bimodals and single-component
clusters is more easily quantified than are small deviations in
ellipticities and core radii between single-component clusters (see
BTa). The efficiency of the PRs at classifying clusters into a broad
range of morphological types is illustrated by their success at
quantitatively discriminating the $ROSAT$ clusters along the lines of
the morphological classes of Jones \& Forman (1992) (see BTb).  There
is a lower frequency of nearly equal-sized bimodals in the $ROSAT$
sample than clusters with more regular morphologies.  Hence, to make
most effective use of the PRs the models need to be adequately sampled
(i.e. simulations have enough clusters) to ensure that relatively rare
regions of PR-space are sufficiently populated.

In this paper we build on the previous studies and investigate the
ability of the PRs to distinguish between models having different
values of $\Omega_0$. Unlike the previous theoretical studies of
cluster morphologies mentioned above, we also consider models having
different power spectra, $P(k)$, since $P(k)$ should affect the
structures of clusters as well.  At the time we began this project it
was too computationally costly to use hydrodynamical simulations to
generate for several cosmological models a large, statistically
robust, number of clusters with sufficient resolution. To satisfy the
above criteria and computational feasibility we instead used pure
N-body simulations.

The organization of the paper is as follows. We discuss the selection
of cosmological models in \S \ref{models}; the specifications of the
N-body simulations in \S \ref{sample}; the validity of using
dark-matter-only simulations to generate X-ray images and the
construction of the images in \S \ref{xray}; and computation of the
PRs in \S \ref{prs}.  We analyze the models having different values of
$\Omega_0$ and a cosmological constant in \S \ref{omega}, and models
with different spectral slopes and $\sigma_8$ in \S \ref{spectrum}.
The implications of the results for all of the models and comparison
of the simulations to the $ROSAT$ sample of BTb is discussed in \S
\ref{disc}.  Finally, in \S \ref{conc} we present our conclusions.

\section{Simulations\label{sim}}

\subsection{Cosmological Models\label{models}}

To test the sensitivity of cluster morphologies to the cosmological
density parameter due to matter, $\Omega_0$, and the power spectrum of
density fluctuations, $P(k)$, we examined several variants of the
standard Cold Dark Matter (CDM) model (e.g., Ostriker 1993).  In Table
\ref{table.models} we list the models and their relevant parameters:
$\Omega_0$; $\lambda_0=\Lambda/3H^2_0$, where $\Lambda$ is a
cosmological constant and $H_0$ is the present value of the Hubble
parameter; the spectral index, $n$, of the scale-free power spectrum
of density fluctuations, $P(k)\propto k^n$; and $\sigma_8$, the
present rms density fluctuations in spheres of radius $8h^{-1}$ Mpc,
where $h$ is defined by $H_0=100h$ km s$^{-1}$ Mpc$^{-1}$.

The parameters of the open CDM model (OCDM) and low-density, flat
model (LCDM) were chosen to be consistent with current observations
(e.g., Ostriker \& Steinhardt 1995).  Their normalizations were set
according to the $\sigma_8 - \Omega_0$ relationship of Eke, Cole, \&
Frenk (1996) to agree with the observed abundance of X-ray clusters.
The biased CDM model (BCDM) was also normalized in this way. However,
the BCDM simulation, because it has $\Omega_0=1$, necessarily has
poorer resolution (i.e. fewer particles per cluster) than the OCDM and
LCDM models due to the fixed box size of our simulations (see \S
\ref{xray}).  For the purposes of our investigation of cluster
morphologies it is paramount to compare simulations having similar
resolution. Hence we use the SCDM model (with $\sigma_8=1$) as our
primary $\Omega_0=1$ simulation for analysis, which has resolution
equivalent to the OCDM and LCDM simulations. (We show in \S \ref{svsb}
that the means of the PR distributions for BCDM and SCDM are very
similar which turns out to be most important for examining the effects
of $\Omega_0$.) Hence, the SCDM, OCDM, and LCDM models allow us to
explore the effects of $\Omega_0$ and $\lambda_0$ on the cluster
morphologies; comparing SCDM and BCDM provides information on the
influence of $\sigma_8$.

We explore the effects of different $P(k)$ on the PRs using the
scale-free models, which have different $n$ from SCDM.  For the
scale-free models we normalized each to the same characteristic mass,
$M_{\star}$, defined to be (Cole \& Lacey 1996) the mass scale when
the linear rms density fluctuation is equal to $\delta_c$, the
critical density for a uniform spherically symmetric perturbation to
collapse to a singularity. For $\Omega_0=1$, the linear theory
predicts $\delta_c \approx 1.686$ (e.g., Padmanabhan 1993).  We take
the SCDM model with $\sigma_8=1$ as a reference for these scale-free
models which gives a characteristic mass of $10^{14}M_{\sun}$. This
procedure allows a consistent means to normalize the scale-free models
relative to each other on the mass scales of clusters. Unfortunately,
as a result of this normalization procedure, at earlier times the
models have different large-scale power and thus the cluster mass
functions are different for each of the models.  The scale-free model
with $n=-1.5$ is similar to the SCDM model and will be used to
``calibrate'' the scale-free models with respect to the other models
(see Table \ref{table.models}).

\subsection{N-body Cluster Sample\label{sample}}

We use the Tree-Particle-Mesh (TPM) N-body code (Xu 1995b) to simulate
the dissipationless formation of structure in a universe filled with
cold dark matter. The simulations consist of $128^3$ particles in a
square box of width $200h^{-1}$ Mpc. The gravitational softening
length is $25h^{-1}$ kpc which translates to a nominal resolution of
$\sim 50h^{-1}$ kpc. This resolution is sufficient for exploring the
structure of clusters with PRs in apertures of radii $R_{ap}\gtrsim
0.5$ Mpc; for a discussion of the related effects of resolution on the
performance of PRs on $ROSAT$ X-ray images see Buote \& Tsai (1995b,
\S 4). All of the realizations have the same initial random phase.

For each simulation we located the 39 most massive clusters using a
version of the DENMAX algorithm (Bertschinger \& Gelb 1991) modified
by Xu (1995a).  This convenient selection criterion yields well
defined samples for each simulation and allows consistent statistical
comparison between different simulations which is the principal goal
of our present investigation. For the various cosmological models we
explore (see Table \ref{table.models}) these clusters generally have
masses ranging from $(0.3-3)\times h^{-1}10^{15}M_{\sun}$, which
correspond to typical cluster masses observed in X-ray (e.g., Edge et
al. 1990; David et. al. 1993) and optically (e.g., Carlberg
et. al. 1995) selected samples.

\subsection{X-ray Images\label{xray}}

\subsubsection{Motivation for $j_{g} \propto \rho^2_{DM}$}

By letting the gas density trace the dark matter density
$(\rho_{gas}\propto\rho_{DM})$ and by assuming that the plasma
emissivity of the gas is constant, we computed the X-ray emissivity of
the clusters, $j_{g} \propto \rho^2_{DM}$.  Given its importance on
the results presented in this paper, here we discuss at some length
the suitability of this approximation.  (Cooling of the gas is
discussed in \S \ref{rosat}.)

For clusters in the process of formation or merging, the gas can have
hot spots appearing where gas is being shock heated (e.g., Frenk,
Evrard, Summers, \& White 1996).  One effect of such temperature
fluctuations on the intrinsic X-ray emissivity is that the intrinsic
plasma emissivity will vary substantially over the cluster thus
rendering $j_{g} \propto \rho^2_{DM}$ a poor approximation.  However,
the intrinsic X-ray emissivity is not observed, but rather that which
is convolved with the spectral response of the detector. For $ROSAT$
observations of clusters with the PSPC, the plasma emissivity is
nearly constant over the relevant ranges of temperatures (NRA
91-OSSA-3, Appendix F, $ROSAT$ Mission Description), and thus
temperature fluctuations contribute negligibly to variations in the
emissivity; for previous discussions of this issue for the PRs see BTa
and TB.

A more serious issue is whether the shocking gas invalidates the
$\rho_{gas}\propto\rho_{DM}$ approximation, in which case the
dynamical state inferred from the gas would not reflect that of the
underlying mass.  TB, who analyzed the hydrodynamical simulation of
Navarro et al. (1995a), showed that the PRs computed for both the gas
and the dark matter gave similar indications of the dynamical states
of the simulated clusters (see \S 4 of TB). In particular, this
applied at early times when the clusters underwent mergers with
massive subclusters. \footnote{See Buote \& Tsai (1995a) for a related
discussion of the evolution of the shape of the gas and dark matter in
the Katz \& White (1993) simulation.}  Hence, our approximation for
the X-ray emissivity should be reasonable even during the early,
formative stages of clusters.

Another possible concern with setting $\rho_{gas}\propto\rho_{DM}$ is
that a gas in hydrostatic equilibrium, which should be a more
appropriate description for clusters in the later stages of their
evolution, traces the shape of the potential of the gravitating matter
which is necessarily rounder than the underlying mass; if the gas is
rounder, then the PRs will be {\it smaller}. However, the core radius
(or scale length) of the radial profile of the gas also influences the
PRs.  In fact, clusters with larger core radii have {\it larger} PRs;
see BTa who computed PRs for toy X-ray cluster models having a variety
of ellipticities and core radii.

When isothermal gas, which is a good approximation for a nearly
relaxed cluster, is added to the potential generated by an average
cluster formed in a $\Omega_0=1$, CDM simulation, the gas necessarily
has a {\it larger} core radius than that of the dark matter (see
Figure 14 of Navarro, Frenk, \& White 1995b). Hence, a gas in
hydrostatic equilibrium will have a larger core radius than that of
the dark matter, at least in the context of the CDM models we are
studying.  Considering the competing effects of smaller ellipticity
and larger core radii (factors of 2-3 in each), and from consulting
Table 6 of BTa, we conclude that no clear bias in the PRs is to be
expected by assuming that the gas follows the dark matter.  In further
support of this conclusion are the similarities of the morphologies of
clusters in the N-body study of Jing et al. (1995). Using
centroid-shifts and axial ratios they find similar results when
$\rho_{gas}\propto\rho_{DM}$ and when the gas is in hydrostatic
equilibrium (see their Figures 5, 6, and 8).

\subsubsection{Construction of the Images}

Having chosen our representation for the X-ray emissivity, we then
generated two-dimensional ``images'' for each cluster.  A rectangular
box of dimensions $4\times 4\times 10$ $h^{-3}$ Mpc$^3$ with random
orientation was constructed about each cluster.  We converted the
particle distribution for each cluster to a mass density field using
the interpolation technique employed in Smoothed Particle
Hydrodynamics (SPH) (e.g., Hernquist \& Katz 1989), from which the
X-ray emissivity was generated, $j_{g} \propto \rho^2_{DM}$.  The SPH
interpolation calculates the density at a grid point by searching for
the nearest neighbors and is thus more robust and physical than other
linear interpolation schemes like Cloud-in-Cell.  For our SPH
interpolation we use 20 neighbors and the spline kernel described in
Hernquist \& Katz.  The interpolation result is independent of the
cell we choose for the X-ray emissivity calculations.

Typically, the boxes contained $\sim 2500$ particles for a cluster;
e.g., SCDM (1799-3965), OCDM (1359-3656), LCDM (1439-3877), and BCDM
(603-1740).  We projected the emissivity along the long edge of the
box into a square $4\times 4$ $h^{-2}$ Mpc$^2$ ``image'' consisting of
``pixels'' of width $20h^{-1}$ kpc. This pixel width was chosen
sufficiently small so as not to inhibit reliable computation the PRs.

We do not add statistical noise or other effects associated with real
observations to the X-ray images since our principal objective is to
examine the intrinsic response of cluster morphologies to different
cosmological parameters. However, the investigation of observational
effects on the PRs by BTa, and the derived error bars on the PRs from
$ROSAT$ clusters by BTb, do not show any large systematic biases;
comparison of the simulations to the $ROSAT$ cluster sample is
discussed in \S \ref{disc}.

\subsection{Power Ratios\label{prs}}

The PRs are derived from the multipole expansion of the
two-dimensional gravitational potential, $\Psi(R,\phi)$, generated by
the mass density, $\Sigma(R,\phi)$, interior to $R$,
\begin{equation}
\Psi(R,\phi) = -2Ga_0\ln\left({1 \over R}\right) -2G
\sum^{\infty}_{m=1} {1\over m R^m}\left(a_m\cos m\phi + b_m\sin
m\phi\right), \label{eqn.multipole}
\end{equation}
where $\phi$ is the azimuthal angle, $G$ is the gravitational constant
and,
\begin{eqnarray}
a_m(R) & = & \int_{R^{\prime}\le R} \Sigma(\vec x^{\prime})
\left(R^{\prime}\right)^m \cos m\phi^{\prime} d^2x^{\prime}\label{eqn.mom1},\\
b_m(R) & = & \int_{R^{\prime}\le R} \Sigma(\vec x^{\prime})
\left(R^{\prime}\right)^m \sin m\phi^{\prime} d^2x^{\prime}\label{eqn.mom2}.
\end{eqnarray}
Because of various advantageous properties of X-ray images of
clusters, we associate the surface mass density, $\Sigma$, with X-ray
surface brightness, $\Sigma_X$ (which is derived from the projection
of $\rho^2_{DM}$ -- see previous section); for more complete
discussions of this association see BTa and TB.  The square of each
term on the right hand side of eq.  (\ref{eqn.multipole}) integrated
over the boundary of a circular aperture of radius $R_{ap}$ is given
by (ignoring factors of $2G$),
\begin{equation}
P_m={1\over 2m^2 R^{2m}_{ap}}\left( a^2_m + b^2_m\right),\label{eqn.powerm}
\end{equation}
for $m>0$ and,
\begin{equation}
P_0=\left[a_0\ln\left(R_{ap}\right)\right]^2,\label{eqn.power0}
\end{equation}
for $m=0$. It is more useful for studies of cluster structure to
consider the ratios of the higher order terms to the monopole term,
$P_m/P_0$, which we call ``power ratios'' (PRs). By dividing each term
by $P_0$ we normalize to the flux within $R_{ap}$. Because for
clusters, $P_m/P_0\ll 1$ for $m>0$ (e.g., BTb), it is preferable to
take the logarithm of the PRs,
\begin{equation}
PR_m \equiv \log_{10}{P_m\over P_0},
\end{equation}
which we shall henceforward analyze in this paper.

Since the $P_m$ depend on the origin of the chosen coordinate system,
we consider two choices for the origin. First, we take the aperture to
lie at the centroid of $\Sigma_X$; i.e. where $P_1$ vanishes. Of these
centroided $PR_m$, $PR_2$, $PR_3$, and $PR_4$ prove to be the most
useful for studying cluster morphologies (see BTa).  In order to
extract information from the dipole term, we also consider the origin
located at the peak of $\Sigma_X$. We denote this dipole ratio by
$P^{(pk)}_1/P^{(pk)}_0$, and its logarithm $PR^{(pk)}_1$, to
distinguish it from the centroided power ratios.

To obtain the centroid of $\Sigma_X$ in a consistent manner for all
clusters we adopted the following procedure. First, when projecting
the cluster (see \S \ref{xray}), the cluster was roughly centered on
the X-ray image by eye.  For each image we computed the centroid in a
circular aperture with $R_{ap}=1.5$ $h^{-1}$ Mpc located about the
field center. This centroid was then used as our initial center for
each cluster; see BTa for a description of how the peak of $\Sigma_X$
is located.

In addition to considering the $PR_m$ individually, we also analyze the
cluster distributions along the ``evolutionary tracks'' in the
$(PR_2, PR_4)$ and $(PR_2,
PR_3)$ planes obeyed by the $ROSAT$ clusters of
BTb.  We refer the reader to TB for a detailed
discussion of the cluster properties along the evolutionary tracks.

Using the augmented Edge et al. (1990) sample of BTb we recomputed the
lines defining the evolutionary tracks of the $ROSAT$ data for the
$1h^{-1}_{80}$ Mpc apertures. This was done since TB selected a subset
of the clusters based on $PR_m$ measurement uncertainty rather than
flux. Following TB we fit $PR_4 = a +
bPR_2$ considering the uncertainties in both axes ; a
similar fit was done for the $(PR_2,
PR_3)$ plane. We obtained $a=-0.92$, $b=1.18$ for the
$(PR_2, PR_4)$ track which we denote by
$PR_{2-4}$. Similarly, for the $(PR_2,
PR_3)$ track, which we denote by $PR_{2-3}$, we obtained
$a=-0.49$, $b=1.16$. These results are nearly the same found by TB for
the slightly different sample.

To facilitate comparison to a previous study of the $PR_m$ of $ROSAT$
clusters (BTb), we compute $PR_m$ of the simulated clusters in apertures
ranging in radius from $0.5h^{-1}_{80}$ Mpc to $1.5h^{-1}_{80}$ Mpc
($H_0=80h_{80}$ km s$^{-1}$ Mpc$^{-1}$) in steps of $0.25h^{-1}_{80}$
Mpc; i.e. $(0.4-1.2)h^{-1}$ Mpc in steps of $0.2h^{-1}$ Mpc.  We refer
the reader to \S 2 of BTa and \S 2 of TB for discussions of the
advantages of using a series of fixed metric aperture sizes to study
cluster morphologies.

\section{$PR_m$ for Models with Different $\Omega_0$ and
$\lambda_0$\label{omega}} 

\begin{figure}
\caption{  \label{fig.scdm} }

\plotone{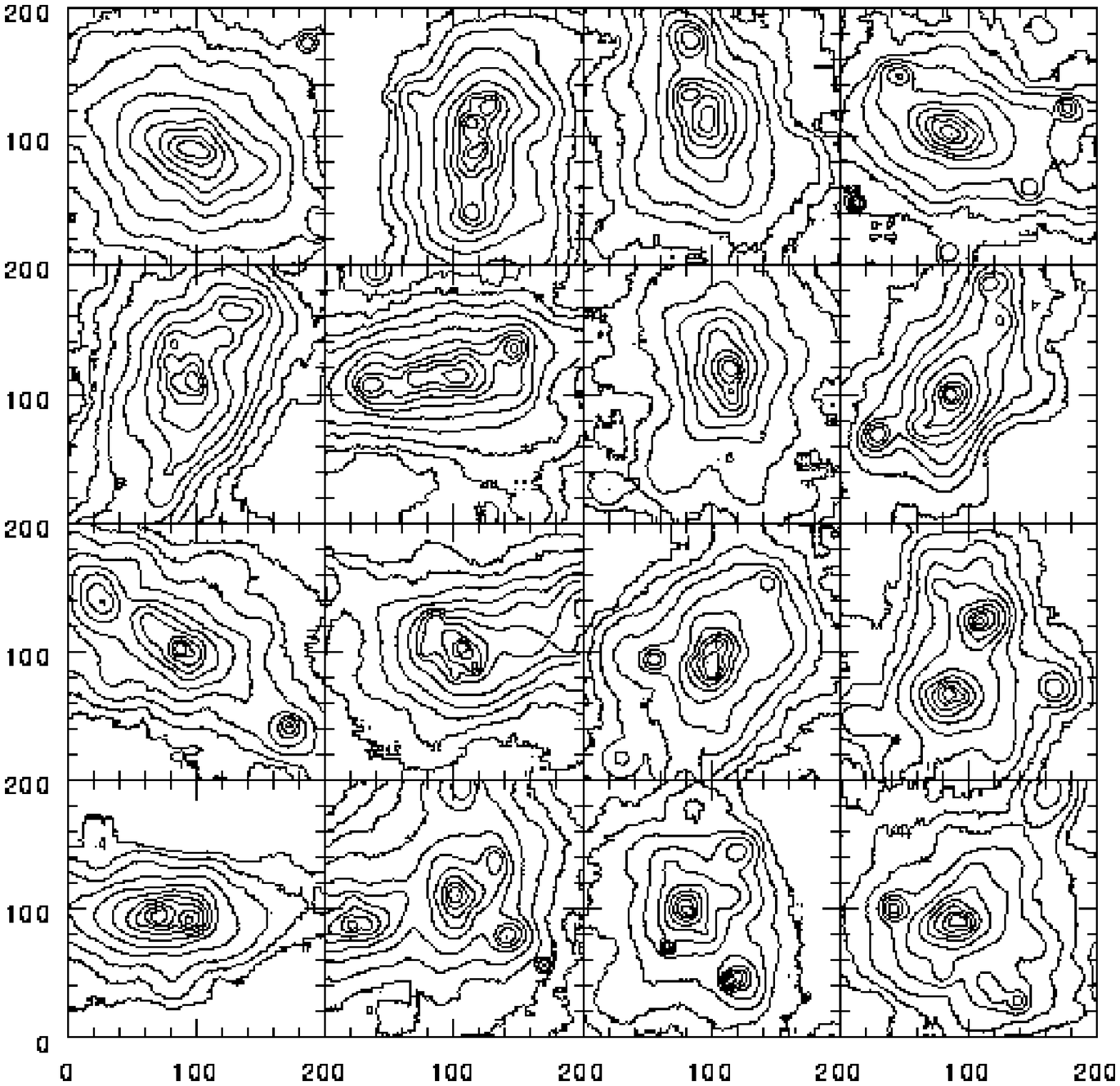}

\raggedright 

Contour plots of the ``X-ray Images'' for 16 of the 39 clusters
analyzed in the SCDM model obtained from projecting $j_{gas}\propto
\rho^2_{DM}$. Each image is $4\times 4$ $h^{-2}$ Mpc$^2$ and the axes
units are $20h^{-1}$ Mpc pixels.

\end{figure}

\begin{figure}
\caption{  \label{fig.ocdm} }

\plotone{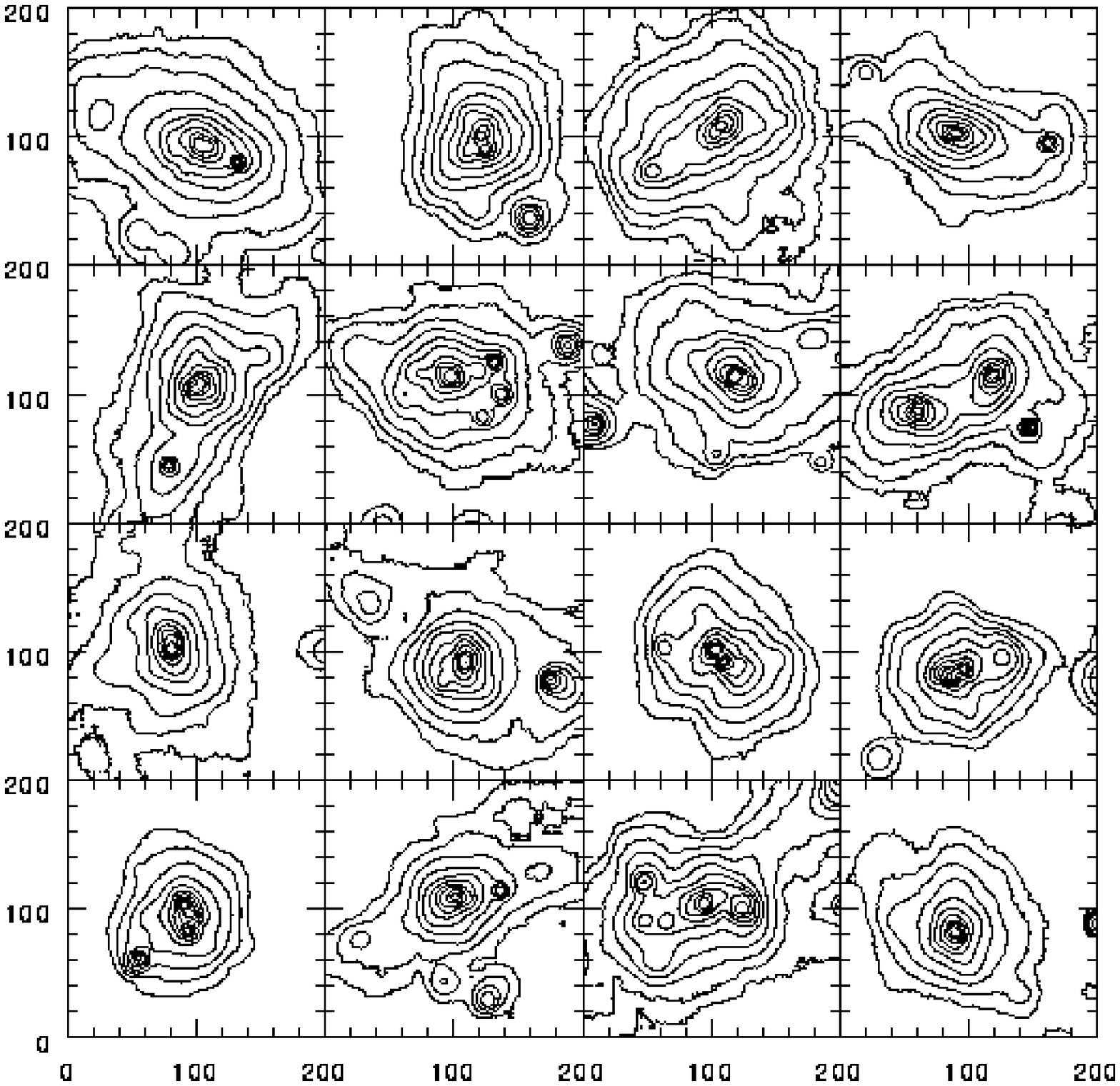}

\raggedright 

As Figure \ref{fig.scdm}, but for OCDM.

\end{figure}

\begin{figure}
\caption{  \label{fig.lcdm} }

\plotone{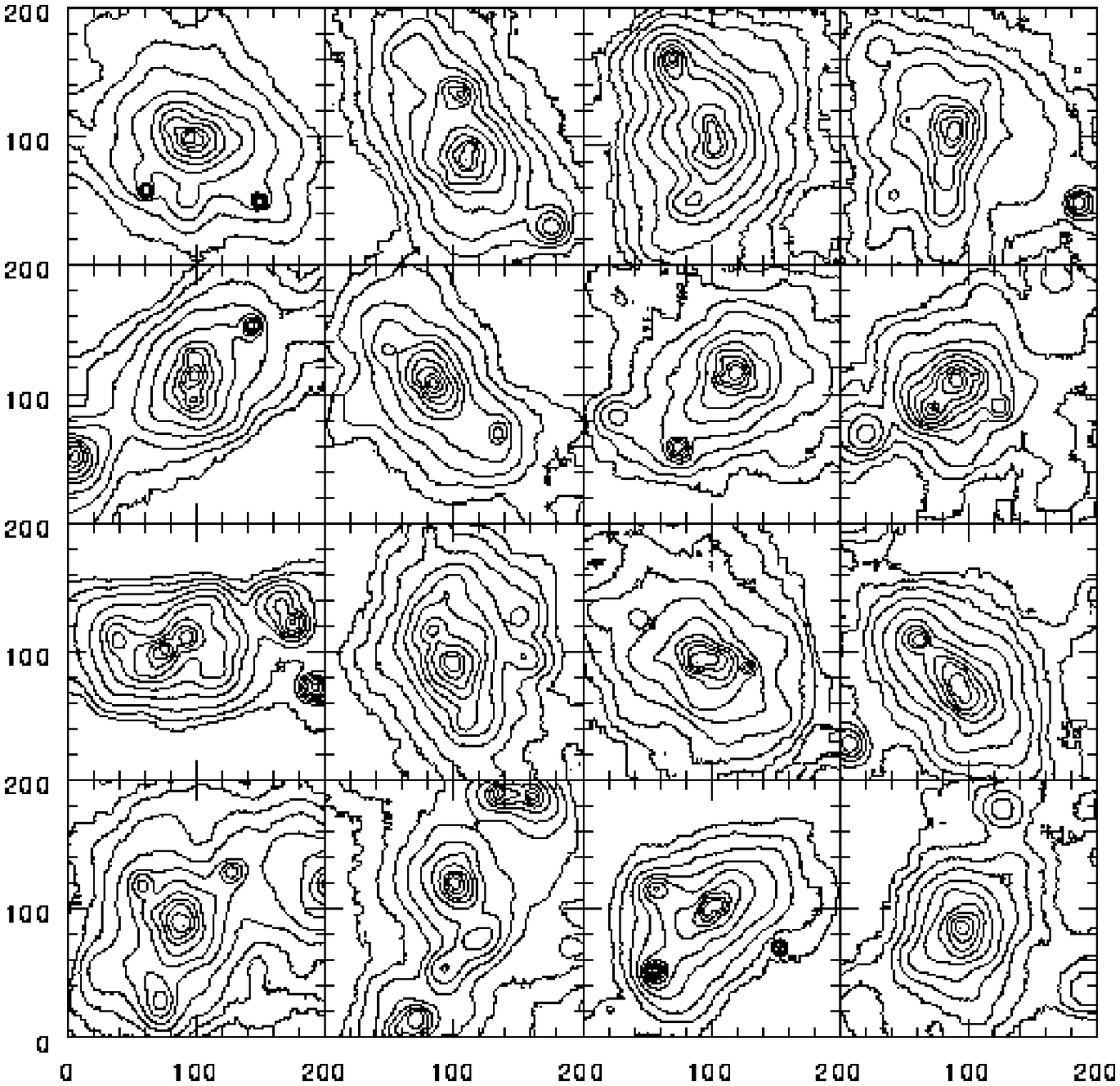}

\raggedright 

As Figure \ref{fig.scdm}, but for LCDM.

\end{figure}

\begin{figure}
\caption{  \label{fig.p2p4cdm} }

\raggedright 

Joint $PR_m$ distributions in the $(0.5,1.0)h^{-1}_{80}$ Mpc apertures
for the SCDM, OCDM, LCDM, and BCDM models.

\plottwo{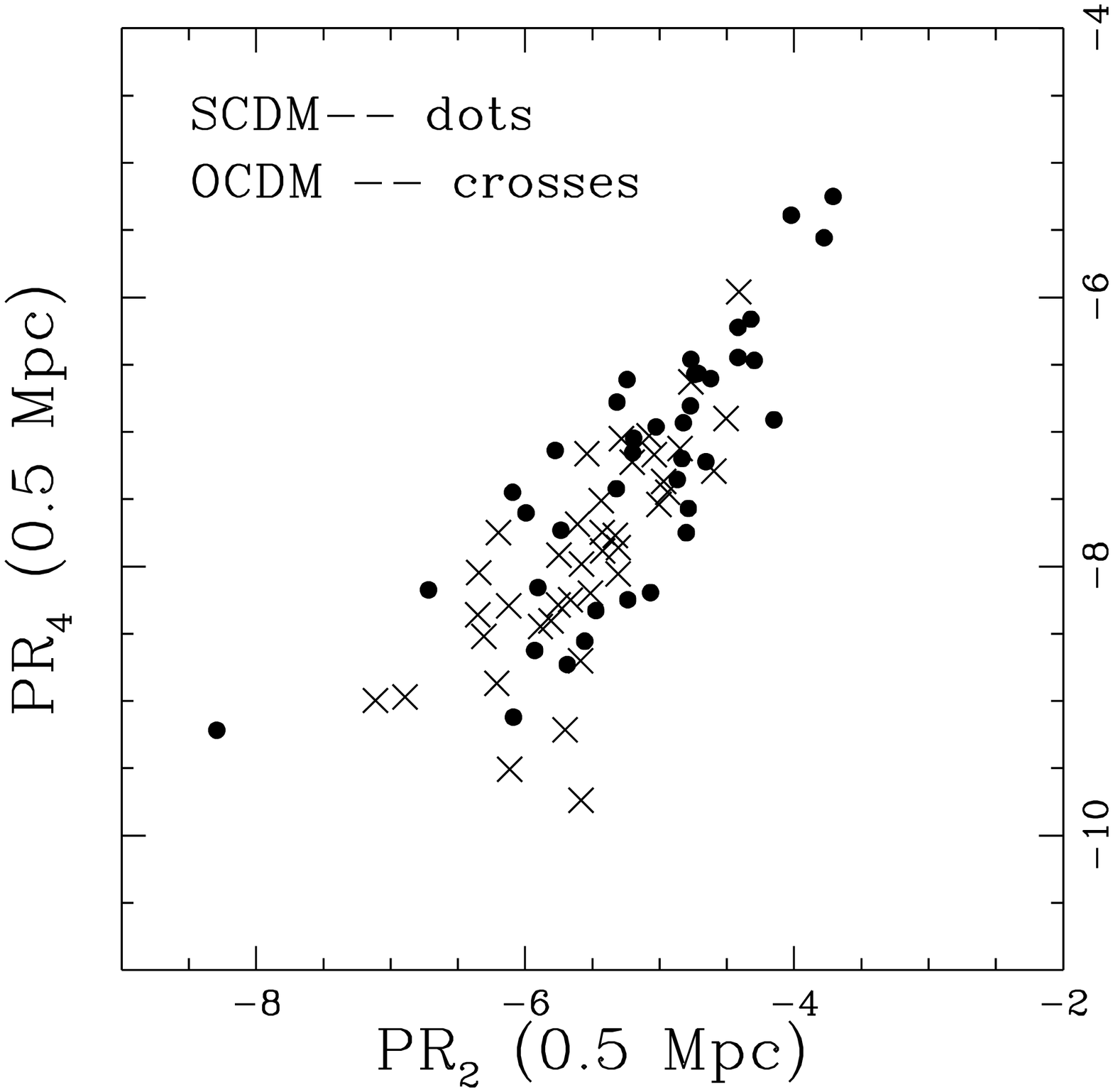}{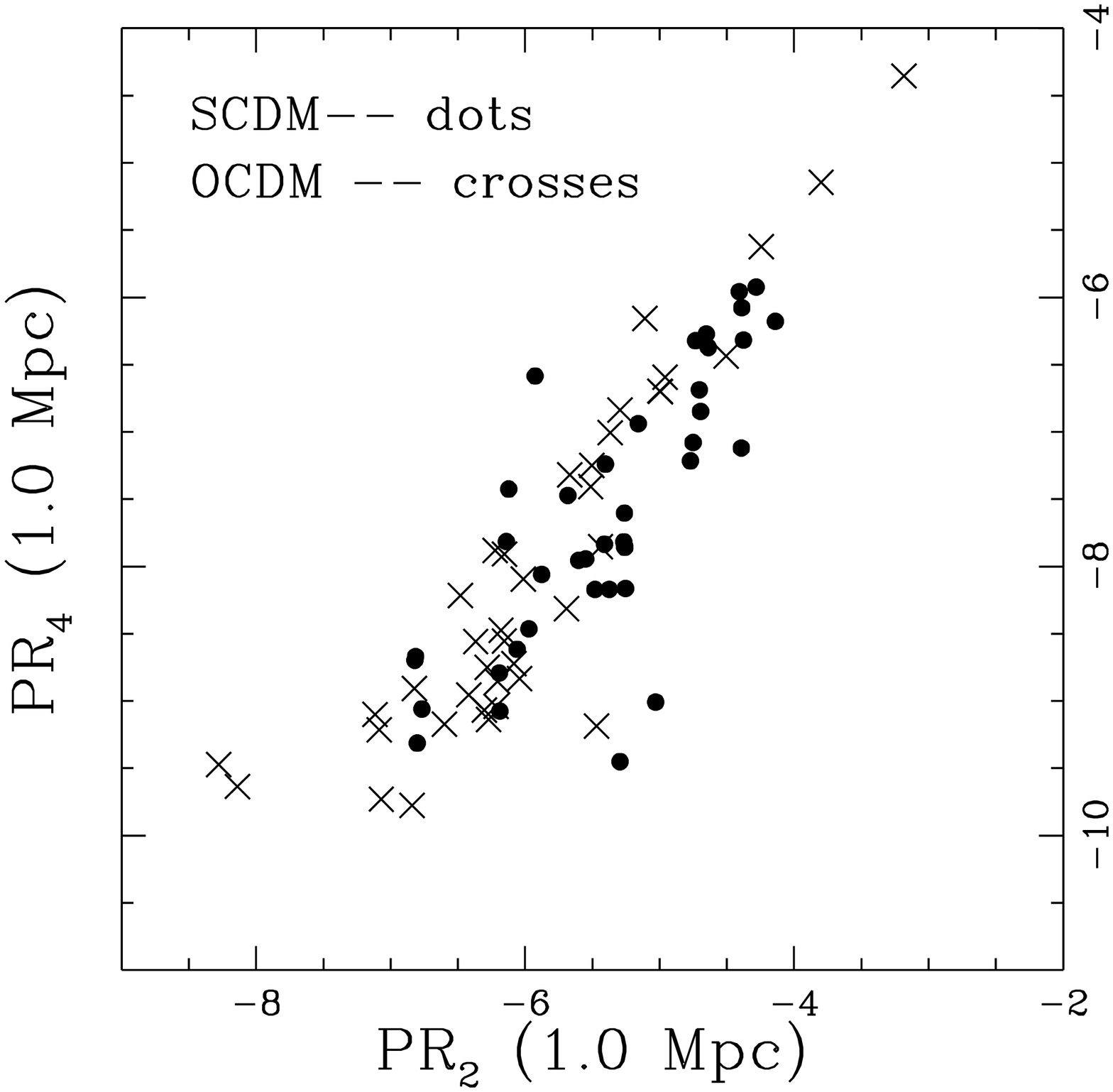}

\plottwo{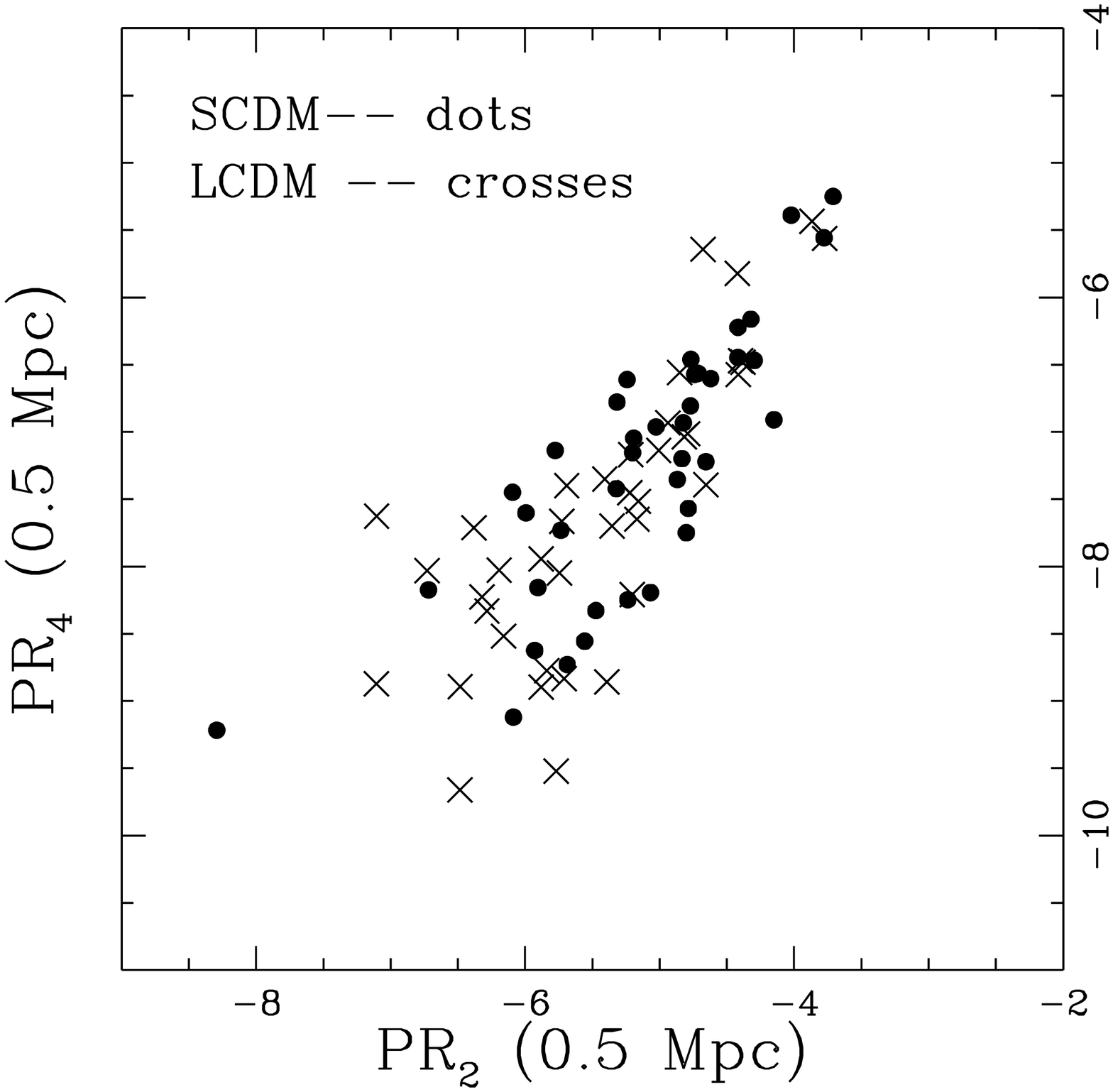}{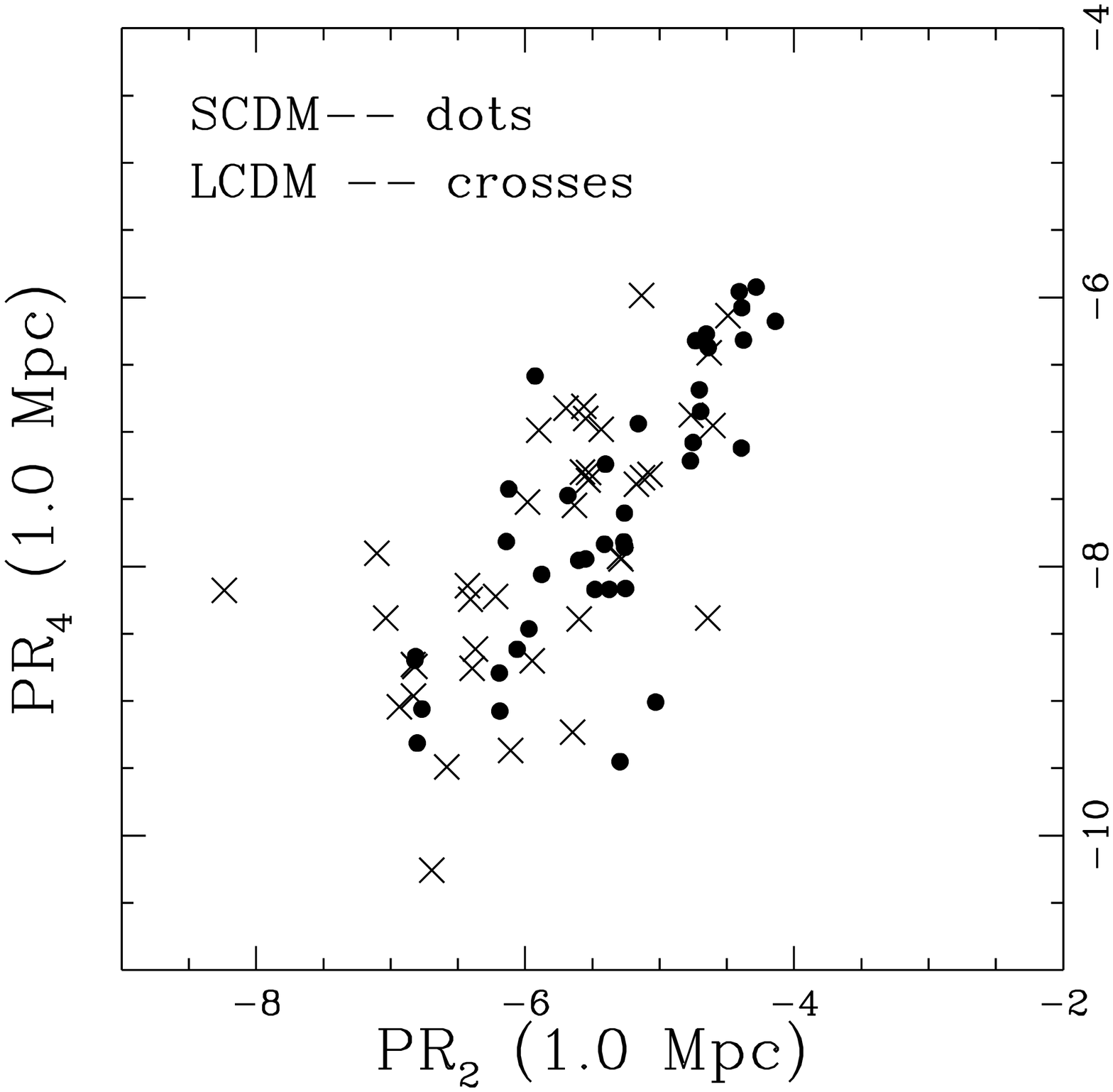}

\plottwo{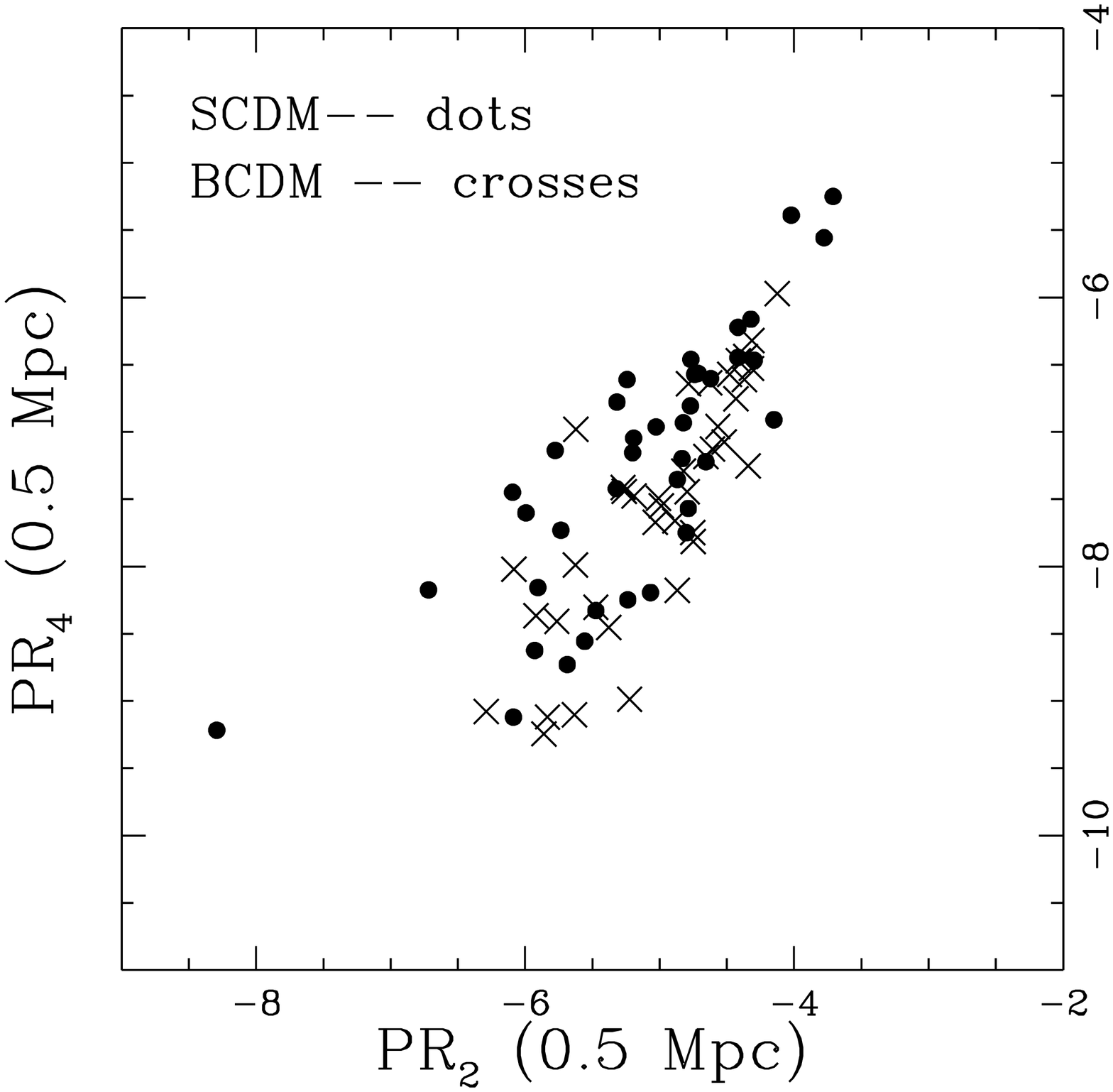}{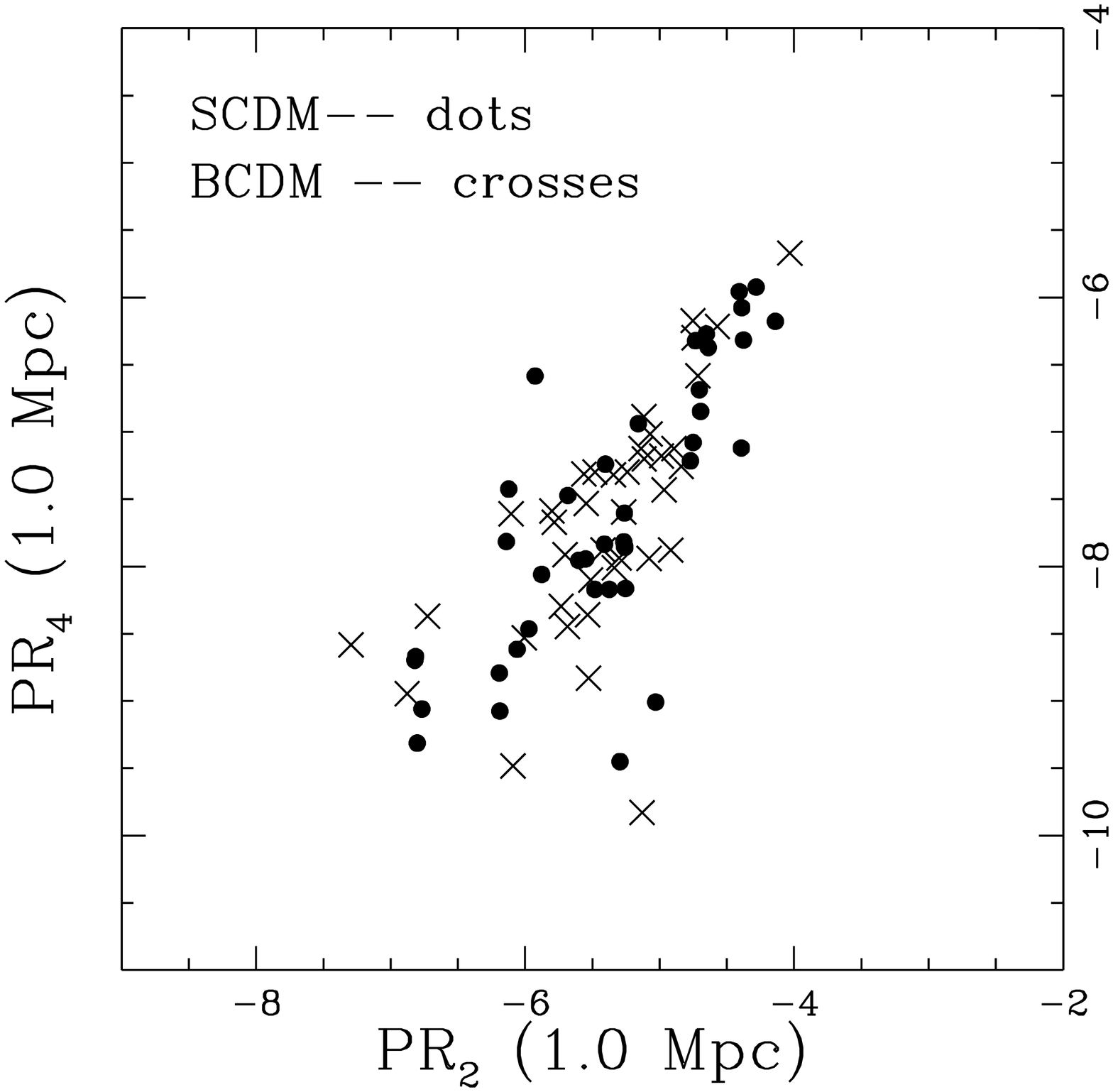}

\end{figure}

First we consider clusters formed in the SCDM, OCDM, and LCDM models.
Contour plots for 16 of the clusters formed in each of the models are
displayed in Figures \ref{fig.scdm}, \ref{fig.ocdm}, and
\ref{fig.lcdm}.  In Figure \ref{fig.p2p4cdm} we show the $(PR_2,
PR_4)$ plane for the $0.5h^{-1}_{80}$ Mpc and $1.0h^{-1}_{80}$ Mpc
apertures; the SCDM model appears in each plot for comparison.  The
clusters in each of the models exhibit tight correlations very similar
to the evolutionary tracks of the $ROSAT$ clusters (BTb) and the
simulated hydrodynamic clusters $(\Omega_0=1)$ of Navarro et
al. (1995a) studied by TB.  Along the evolutionary tracks a shift in
the means of the $PR_m$ is easily noticeable in the $0.5h^{-1}_{80}$
Mpc aperture, being most apparent for the SCDM-OCDM models.  The
spread of the $PR_m$ along the track in the $1.0h^{-1}_{80}$ Mpc
aperture for SCDM-OCDM also appears to be different.  The
distributions perpendicular to $PR_{2-4}$ do not show discrepancies
obvious to the eye.

\begin{figure}
\caption{  \label{fig.prhist.05mpc} }

\raggedright Histograms for the PRs in the $0.5h^{-1}_{80}$ Mpc
aperture. SCDM is given by the solid line, OCDM by the dotted line,
and LCDM by the dashed line.

\plottwo{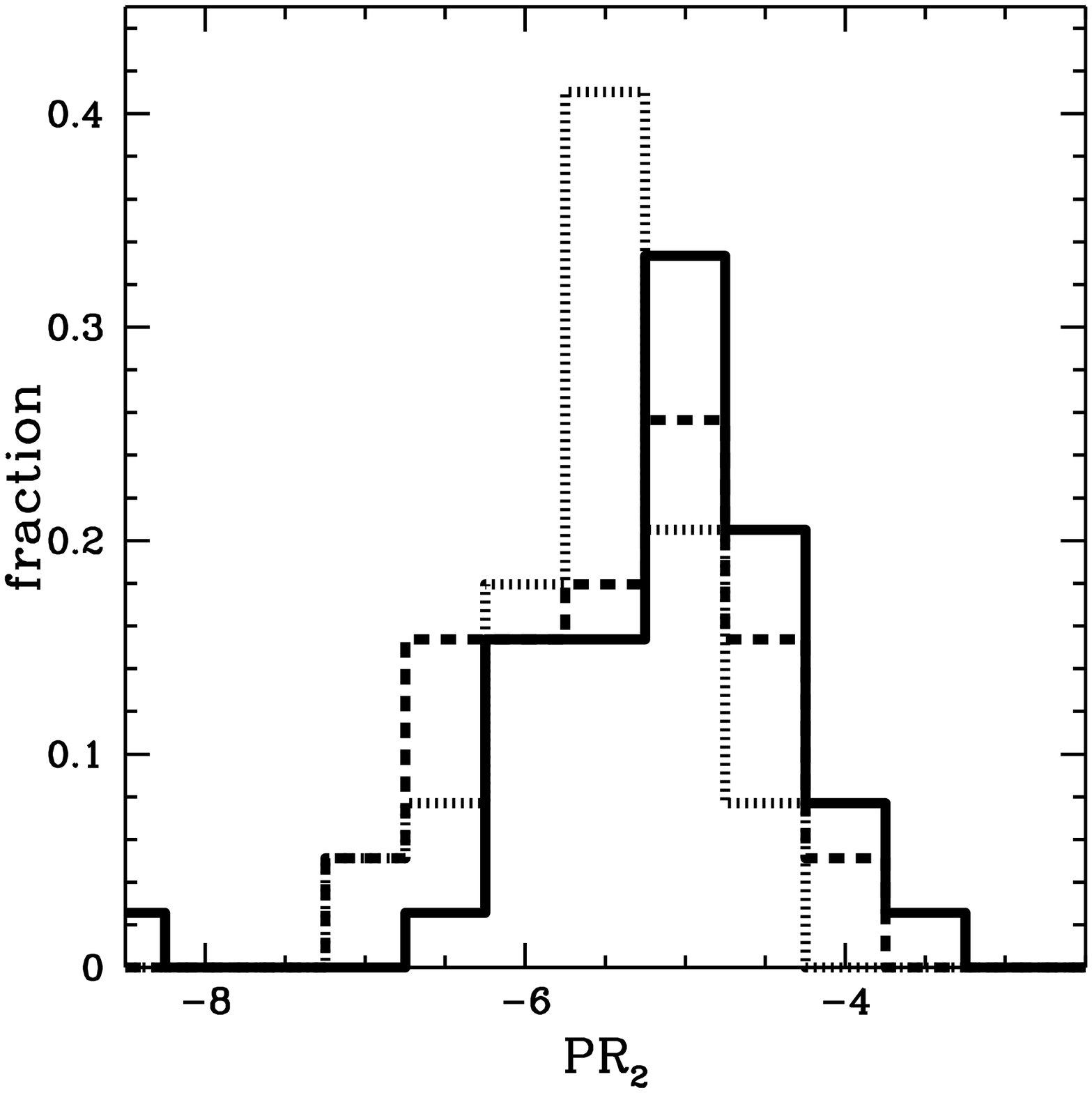}{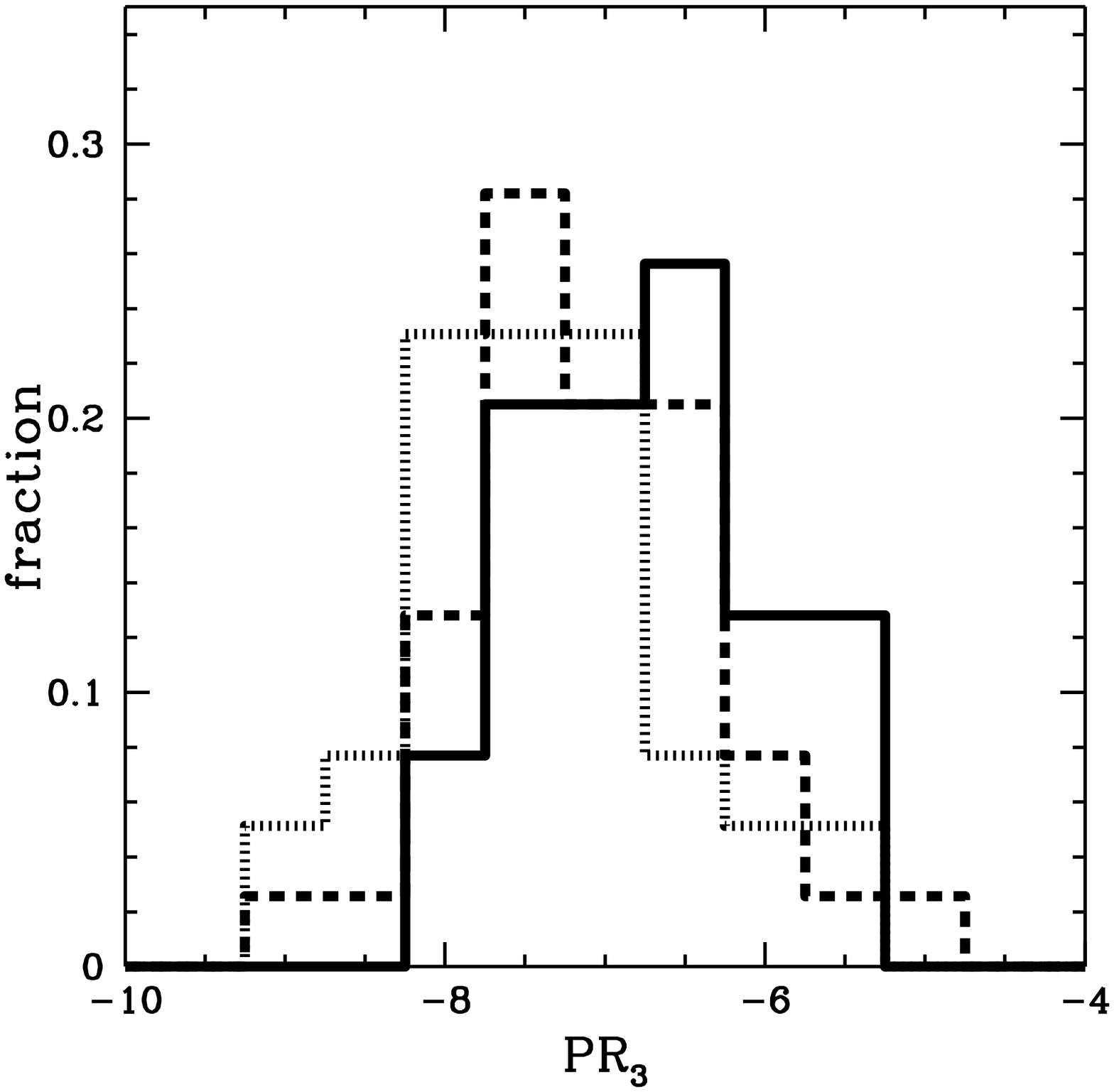}

\end{figure}

\begin{figure}
\caption{  \label{fig.prhist.10mpc} }

\raggedright 

As Figure \ref{fig.prhist.05mpc}, but for the $1.0h^{-1}_{80}$ Mpc
aperture.

\plottwo{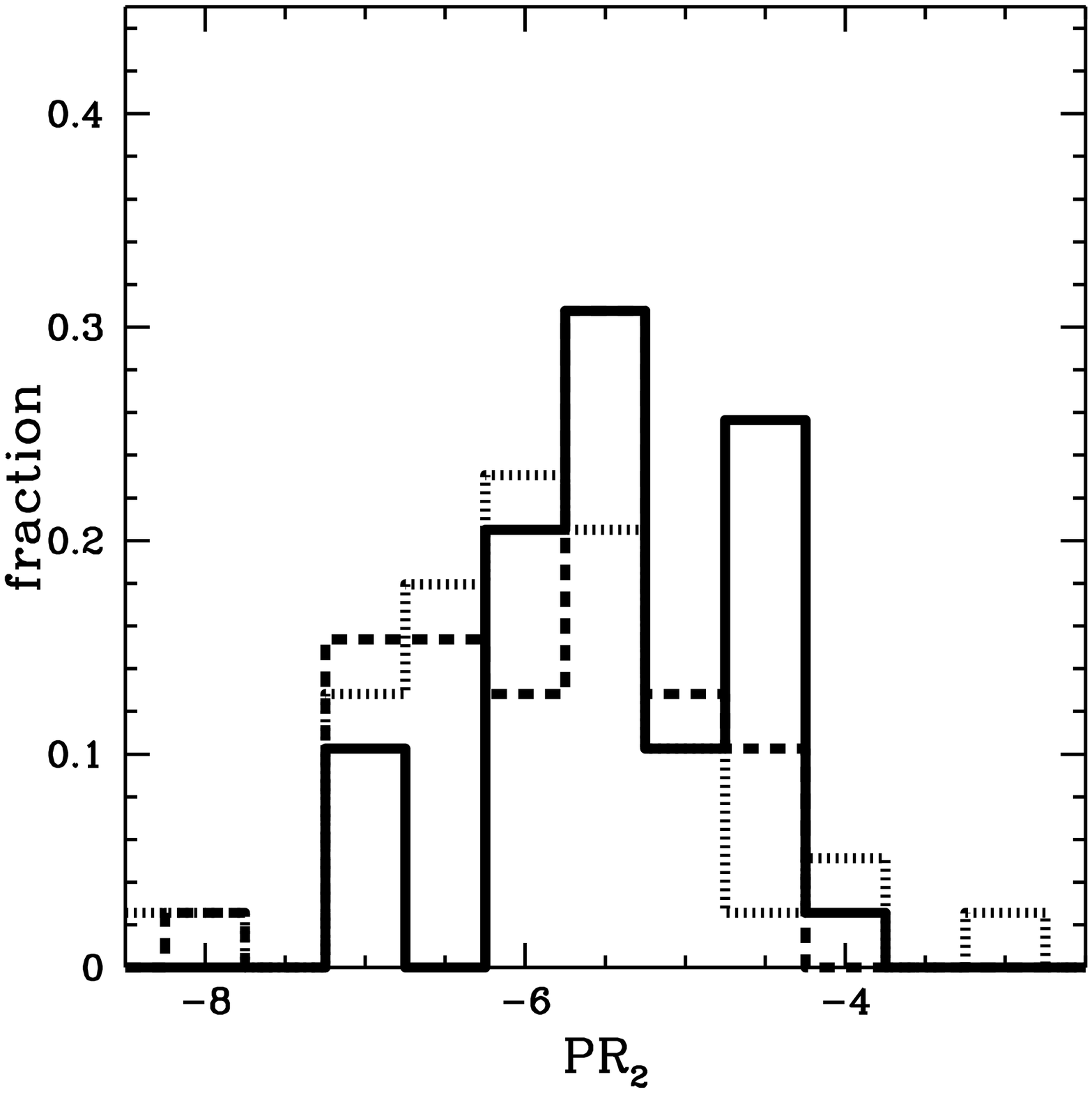}{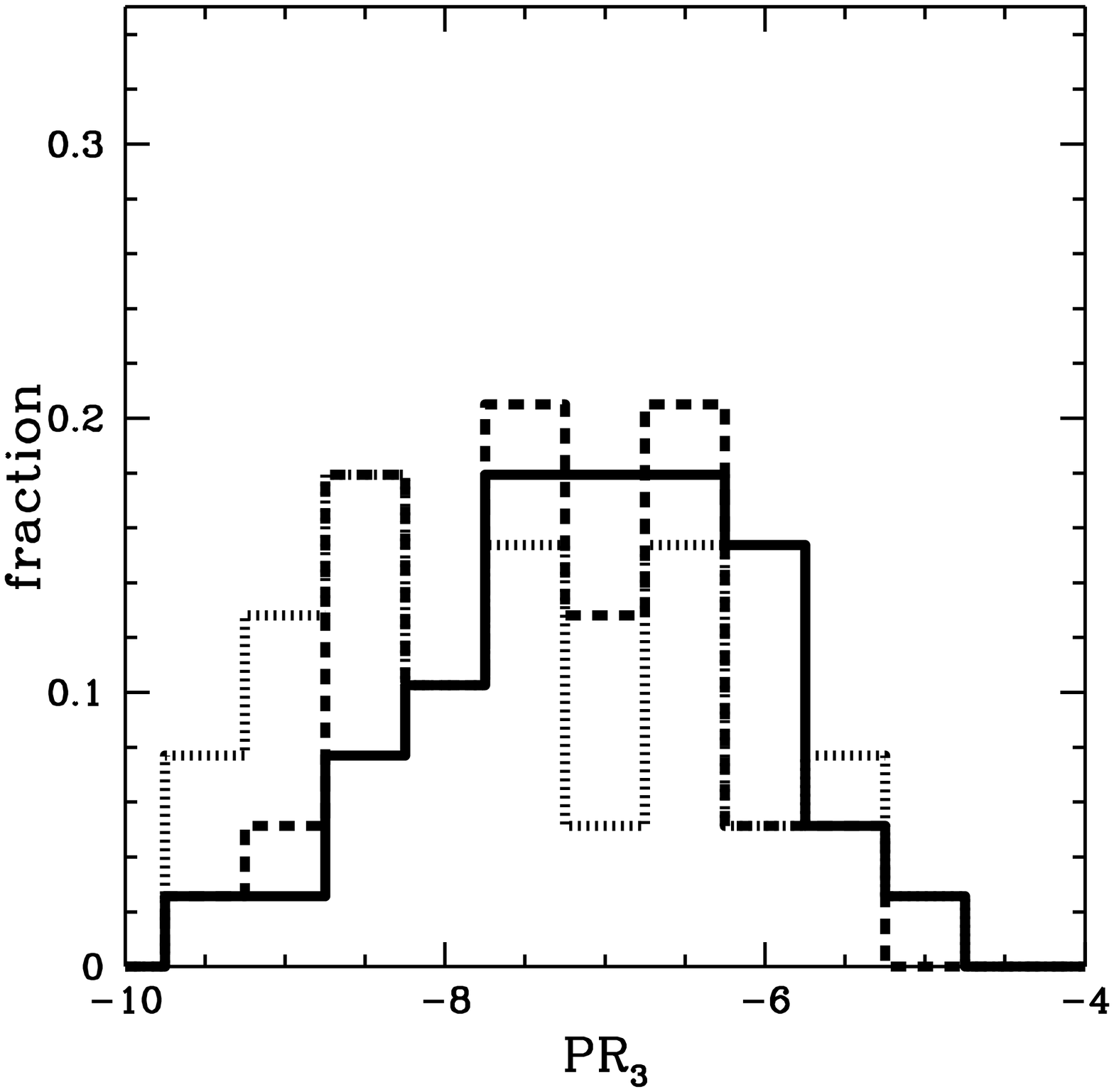}

\end{figure}

At this time we shift our focus away from the evolutionary tracks and
instead analyze the individual $PR_m$ distributions, which prove to be
more powerful for distinguishing between cosmological models as we
show below. We give the individual $PR_m$ distributions of clusters in the
three models for the $0.5h^{-1}_{80}$ Mpc and $1.0h^{-1}_{80}$ Mpc
apertures in Figures \ref{fig.prhist.05mpc} and
\ref{fig.prhist.10mpc}.  We found it most useful to compare these
distributions in terms of their means, variances, and
Kolmogorov-Smirnov (KS) statistics. For the number of clusters in each
of our simulations (39) higher order statistics like the skewness and
kurtosis are unreliable ``high variance'' distribution shape estimates
(e.g., Bird \& Beers 1993).  We did consider more robust statistics
like the ``Asymmetry Index'' (AI), which measures a quantity similar
to the skewness, and the ``Tail Index'' (TI), which is similar to the
Kurtosis (Bird \& Beers 1993). However, we found that they did not
clearly provide useful information in addition to the lower order
statistics and KS test, and thus we do not discuss them
further. \footnote{Actually, the KS test turns out not to provide much
additional information over the t-test and F-test, but we include it
for ease of comparison to previous studies; e.g., Jing et al. (1995);
Mohr et al. (1995); TB.}

\begin{figure}
\caption{  \label{fig.avgpr.omega} }

\raggedright The standard deviation as a function of the average value
of the PRs in the $0.75h^{-1}_{80}$ Mpc aperture for the models in \S
\ref{omega}. The error bars represent $1\sigma$ errors estimated from
1000 bootstrap resamplings.

\plottwo{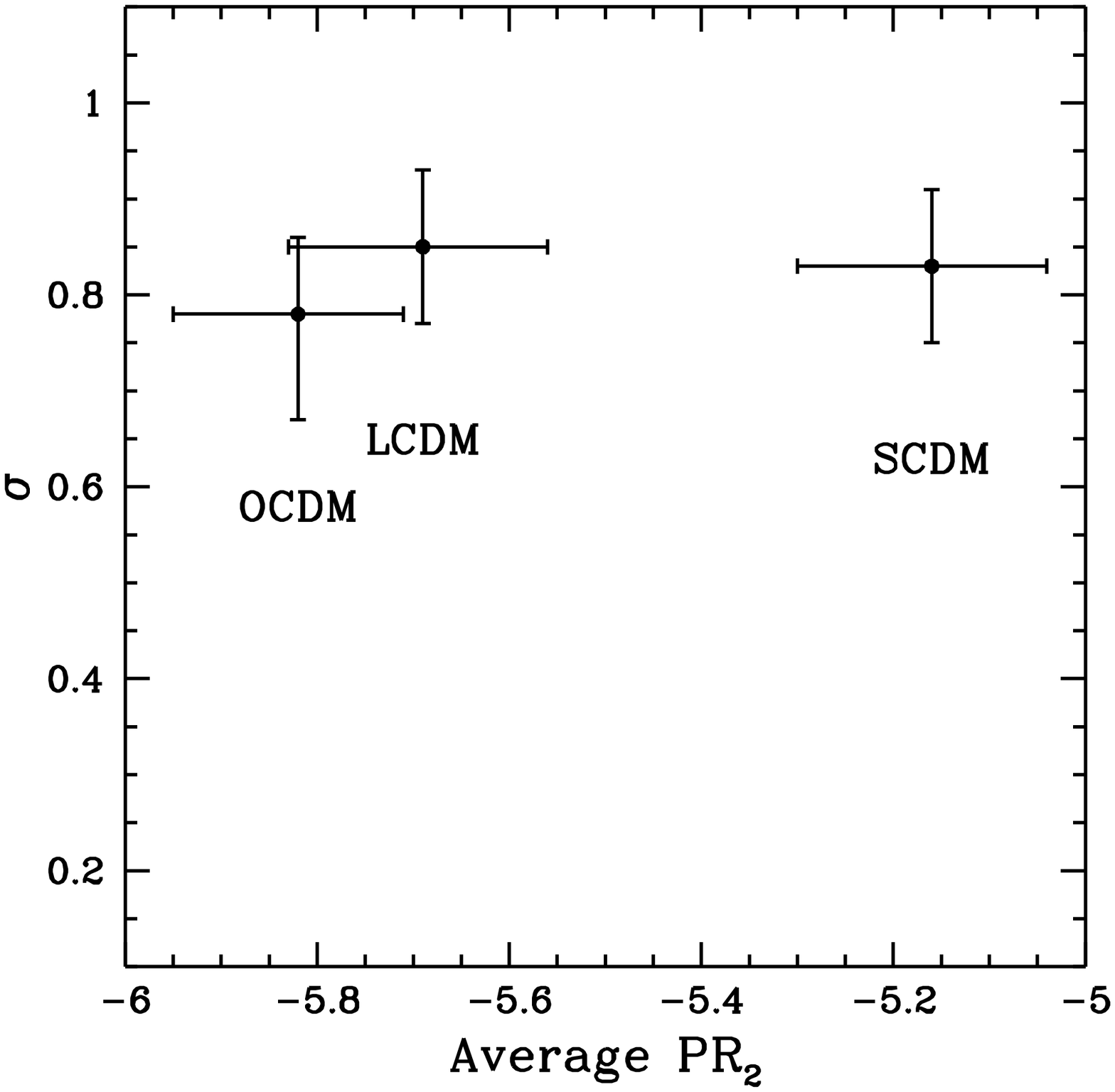}{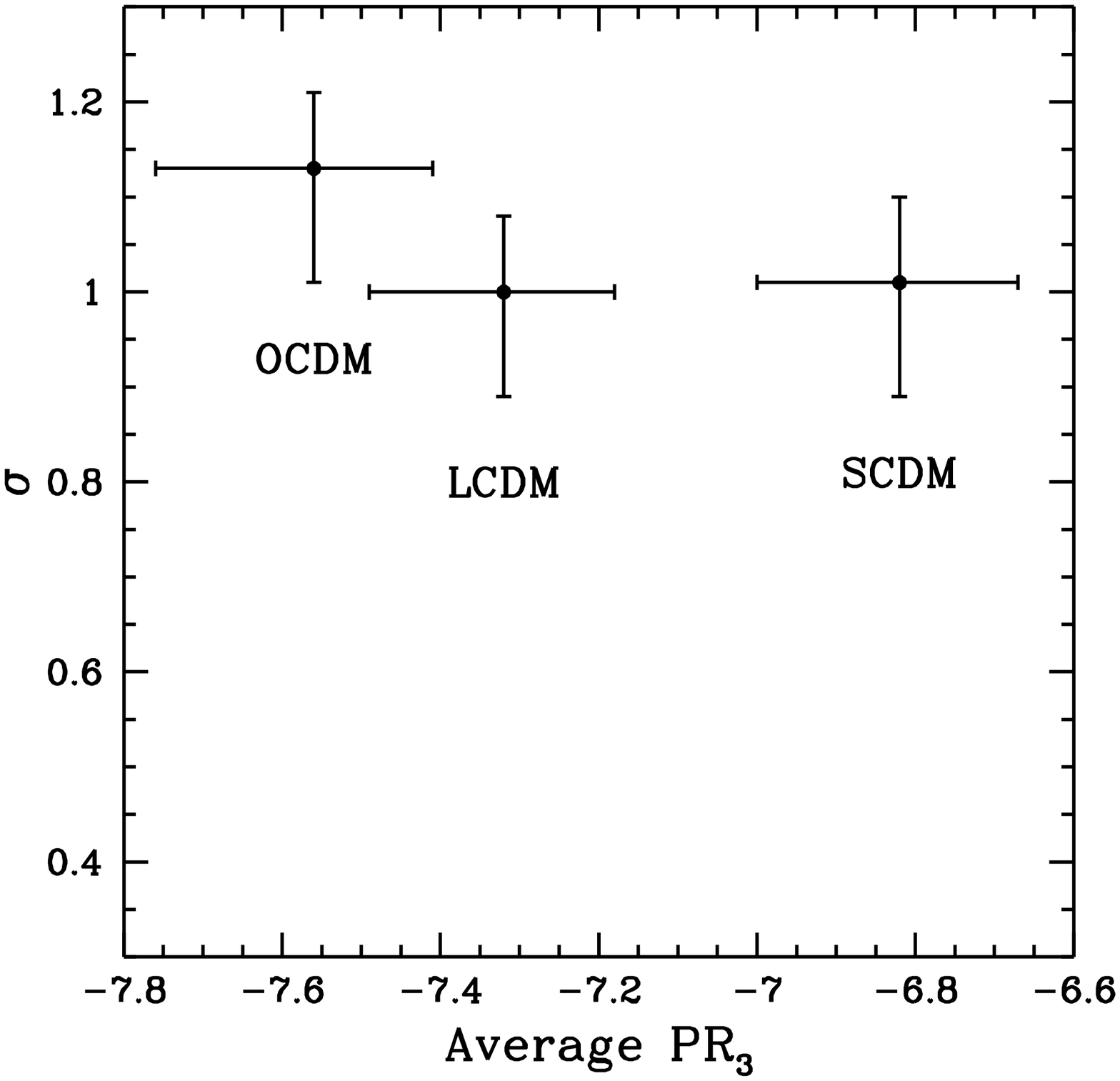}

\end{figure}

The means and standard deviations for the $(0.5,0.75,1.0)h^{-1}_{80}$
Mpc apertures are listed in Table \ref{table.avgpr}.  As a possible
aid to understanding the relationships between the values in Table
\ref{fig.avgpr.omega}, we plot in Figure \ref{fig.avgpr.omega} the
standard deviation vs the mean for the $0.75h^{-1}_{80}$ Mpc aperture.
We do not present the results for the larger apertures because they
did not significantly improve the ability to distinguish between the
models. Moreover, we found that $PR_4$ and $PR^{(pk)}_1$ do not
provide much useful information in addition to $PR_2$ and
$PR_3$. Generally $PR_4$ tracks the behavior of $PR_2$, though showing
less power to discriminate between models; the similarity to $PR_2$ is
understandable given the strong correlation shown in Figure
\ref{fig.p2p4cdm}. Likewise, $PR^{(pk)}_1$ is similar to, but not
quite so effective as, $PR_3$. For compactness, thus, we shall
henceforward mostly restrict our discussion to results for $PR_2$ and
$PR_3$ in the $(0.5,0.75,1.0)h^{-1}_{80}$ Mpc apertures.

We compare the means, standard deviations, and total distributions of
the models in Table \ref{table.test} using standard non-parametric
tests as described in Press et al. (1994).  The Student's t-test,
which compares the means of two distributions, computes a value,
$p_t$, indicating the probability that the distributions have
significantly different means. Similarly, the F-test, which compares
the variances of two distributions, computes a value, $p_F$,
indicating the probability that the distributions have significantly
different variances. Finally, the KS test, which compares the overall
shape of two distributions, computes a value, $p_{KS}$, indicating the
probability that the distributions originate from the same parent
population; the probabilities listed in Table \ref{table.test} are
given as percents; i.e. decimal probability times 100.  Note that for
the cases where the F-test gives a probability less than 5\% we use
the variant of the t-test appropriate for distributions with
significantly different variances (i.e. program {\it tutest} in Press
et. al.).

\subsection{SCDM vs. OCDM}

As is clear from inspection of Figures \ref{fig.p2p4cdm} -
\ref{fig.avgpr.omega}, and Tables \ref{table.avgpr} -
\ref{table.test}, the means of the $PR_m$ of the SCDM model exceed
those of OCDM. In terms of the t-test the significance of the
differences is very high. Of all the $PR_m$, generally the means of
$PR_2$ and $PR_3$ exhibit the largest significant differences; the
most significant differences are seen for $PR_3$ in the
$0.5h^{-1}_{80}$ Mpc aperture, $p_t=0.02\%$, and for $PR_2$ in the
$0.75h^{-1}_{80}$ Mpc aperture, $p_t=0.06\%$.  Hence, though different
in all the apertures, the discrepancy in the means is most significant
for the smallest apertures, $(0.5,0.75)h^{-1}_{80}$ Mpc.

The variances of $PR_3$ in the SCDM model are essentially
consistent at all radii with their corresponding values in
OCDM. However, for $PR_2$ the variances are consistent at
$0.75h^{-1}_{80}$ Mpc, but marginally inconsistent at
$(0.5,1.0)h^{-1}_{80}$ Mpc (and inconsistent at
$(1.25,1.5)h^{-1}_{80}$ Mpc).

The KS test generally indicates a significant difference in SCDM and
OCDM when also indicated by the t-test, or the t-test and F-test
together. The level of discrepancy is usually not as significant as
given by the t-test, except when $p_F$ is small as well. Since the KS
test does not indicate discrepancy when both the t-test and F-test
indicate similarity, we conclude that higher order properties of the
PR distributions are probably not very important for the SCDM and OCDM
models (at least for the samples of 39 clusters in our
simulations). Since this qualitative behavior holds for the other
model comparisons, we shall not emphasize the KS tests henceforward.

Finally, in terms of the various significance tests we find that the
$PR_{2-4}$ distribution essentially gives a weighted probability of the
individual $PR_2$ and $PR_4$ distributions; i.e. it does not enhance
the discrepancy in the individual distributions. Perpendicular to
$PR_{2-4}$ the distributions are consistent. The same behavior is seen
for $PR_{2-3}$ as well. This behavior is seen for the remaining model
comparisons in this section so we will not discuss the joint
distributions further.

\subsection{SCDM vs. LCDM}

The $PR_m$ means for the SCDM clusters also systematically exceed those in
the LCDM model, however the significance of the difference is not as
large as with the OCDM clusters.  The largest discrepancy is observed
for $PR_2$ in the $(0.75,1.0)h^{-1}_{80}$ Mpc apertures for which
$p_t=(0.7\%,0.8\%)$.  The other $PR_m$ show only a marginal discrepancy
in the means. For apertures $(0.5,0.75)h^{-1}_{80}$ Mpc, $PR_3$ has
$p_t=(4\%,3\%)$, but is quite consistent at larger radii.  The
variances for the SCDM and LCDM models are consistent for essentially
all radii and all $PR_m$.

\subsection{OCDM vs. LCDM}

The $PR_m$ means for the LCDM clusters appear to systematically exceed
those in the OCDM model, however the formal significances of the
differences are quite low. The means are entirely consistent at all
radii for $PR_2$.  However, $PR_3$ shows a marginal difference
in the $0.5h^{-1}_{80}$ Mpc aperture ($p_t=9\%$).  The variances of the
$PR_m$ of the OCDM and LCDM models behave similarly as with the SCDM and
OCDM comparison above, as expected since the SCDM-LCDM variances are
essentially identical. However, the degree of discrepancy is not as
pronounced.

\subsection{Performance Evaluation I.}

The means of the individual $PR_m$ distributions generally exhibit the
most significant differences between the SCDM, OCDM, and LCDM models;
the variances are much less sensitive to the models, with $PR_3$
showing no significant variance differences.  The larger means for the
$PR_m$ in the SCDM models are expected from the arguments of, e.g.,
Richstone et al. (1992). That is, in a sub-critical universe the
growth of density fluctuations ceased at an early epoch and so
present-day clusters should show less ``substructure'' than in an
$\Omega_0=1$ universe where formation continues to the
present. Clusters with more structure will have systematically larger
values of the $PR_m$.

The $PR_m$ whose means show the most significant differences between the
models are $PR_2$ and $PR_3$, where $PR_2$ typically performs
best for apertures $(0.75,1.0)h^{-1}_{80}$ Mpc and $PR_3$ is most
effective for $(0.5,0.75)h^{-1}_{80}$ Mpc.  Although useful,
$PR^{(pk)}_1$ is often the least effective $PR_m$ for
differentiating models in terms of its mean; this relatively weak
performance of the dipole ratio with respect to other moments is
echoed in the results of Jing et al. (1995) who found that their
measure of an axial ratio performed better than a centroid shift for
discriminating between models (see their tables 3-6).

\section{$PR_m$ for Models with Different $n$ and
$\sigma_8$\label{spectrum}} 

In this section we investigate the effects of different power spectra
for models otherwise conforming to the specifications of the SCDM
model.  First, we examine models with different spectral indices of
the scale-free power spectrum $(P(k)\propto k^n)$, $n=0,-1,-1.5,-2$.
Then we examine the BCDM model which has a lower power-spectrum
normalization as expressed by $\sigma_8$.  As in the previous section,
we find the $(0.5,0.75,1.0)h^{-1}_{80}$ Mpc apertures to be more
useful than the larger apertures, and that $PR_4$ and $PR^{(pk)}_1$ do
not provide much useful information in addition to that provided by
$PR_2$ and $PR_3$. Hence, for compactness we again mostly restrict the
discussion to $PR_2$ and $PR_3$ in the smaller apertures.

\subsection{$n=-1.5$ vs. SCDM\label{n15}}

Before analyzing the $PR_m$ of models with different $n$ we calibrate
the scale-free models by comparing the $n=-1.5$ scale-free model to
the SCDM model since they should have similar properties (see \S
\ref{models}). We find that the means, variances, and KS statistics of
the centroided $PR_m$ for the SCDM and $n=-1.5$ models are entirely
consistent for all aperture radii with only one possible exception.
The variances of $PR_3$ exhibit a marginal $(p_F=5\%)$ discrepancy
in the $0.5h^{-1}_{80}$ Mpc aperture.  The significance of this
variance discrepancy should be treated with caution given the complete
consistency of the means $(p_t=31\%)$ and KS (32\%) test at this
radius as well as the consistency of all the tests at all the other
radii investigated. Hence, the cluster morphologies of the SCDM and
$n=-1.5$ models are very consistent expressed in terms of the
centroided $PR_m$ ($m=2,3,4$).

\subsection{$n=0$ vs. $n=-2$\label{n0n2}}

\begin{figure}
\caption{  \label{fig.ets.sf} }

\raggedright 

Joint $PR_m$ distributions in the $(0.5,1.0)h^{-1}_{80}$ Mpc apertures
for the scale-free models: SF00 $(n=0)$ denoted by dots and SF20
$(-2)$ denoted by crosses.

\plottwo{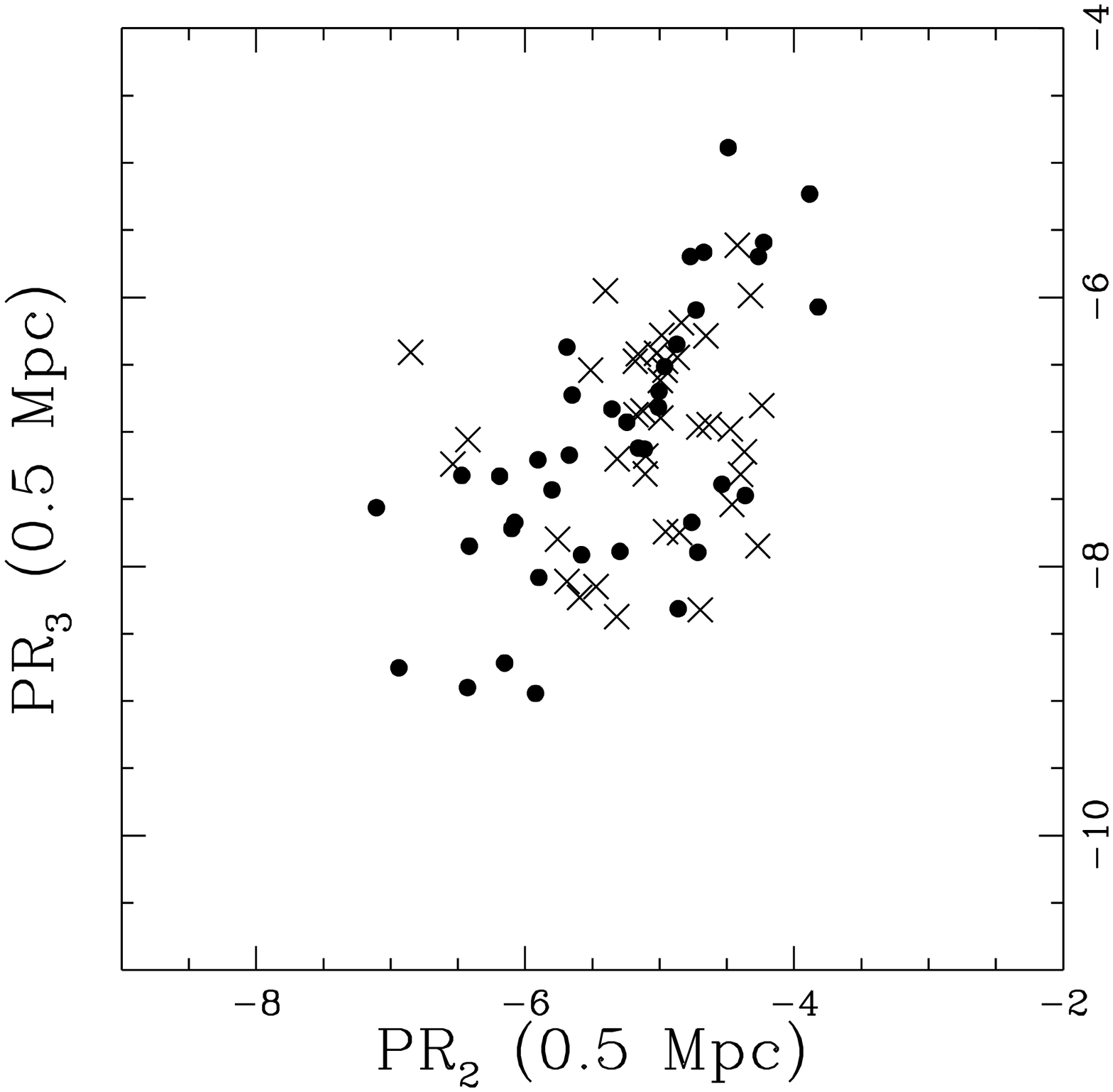}{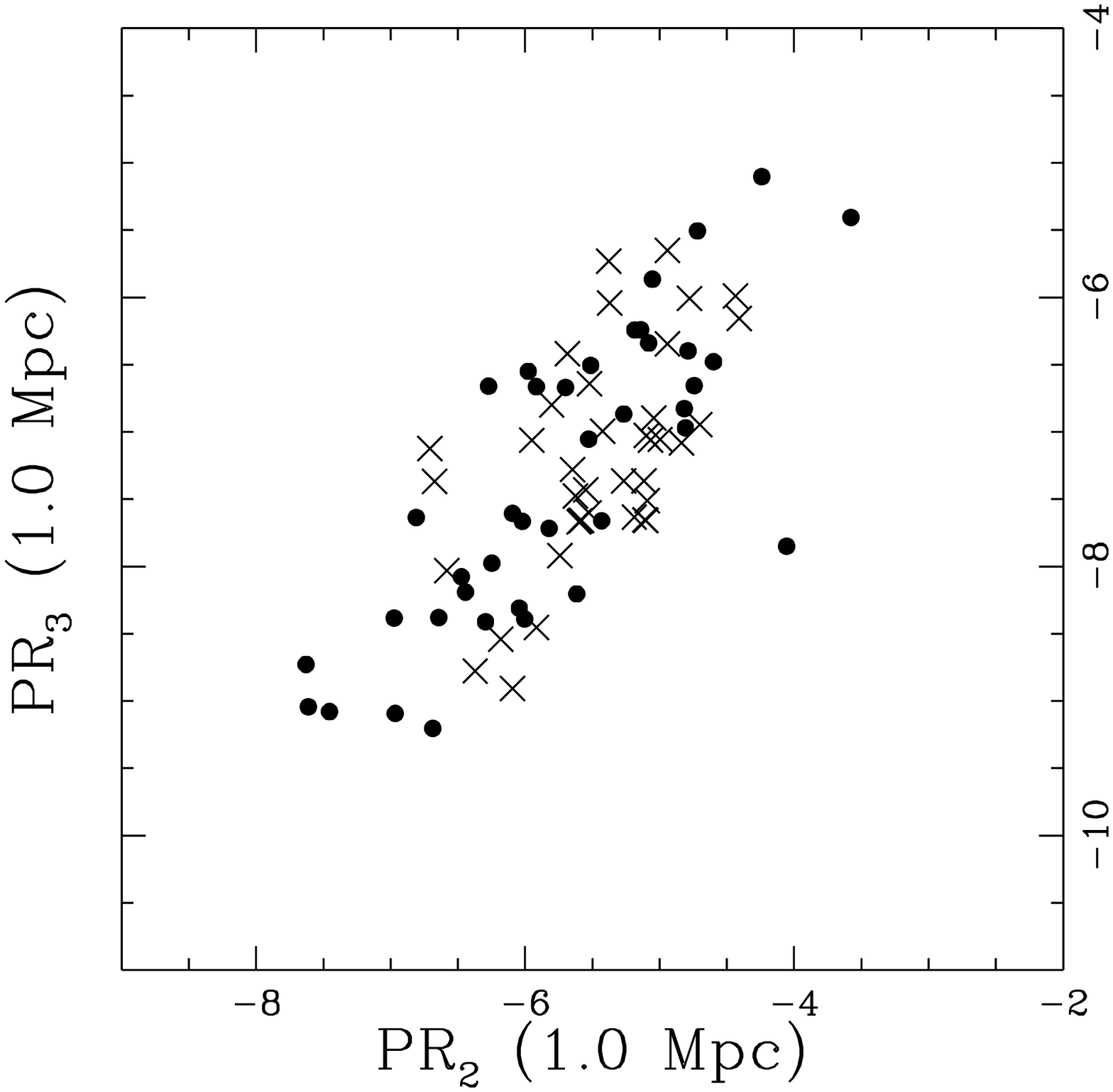}

\plottwo{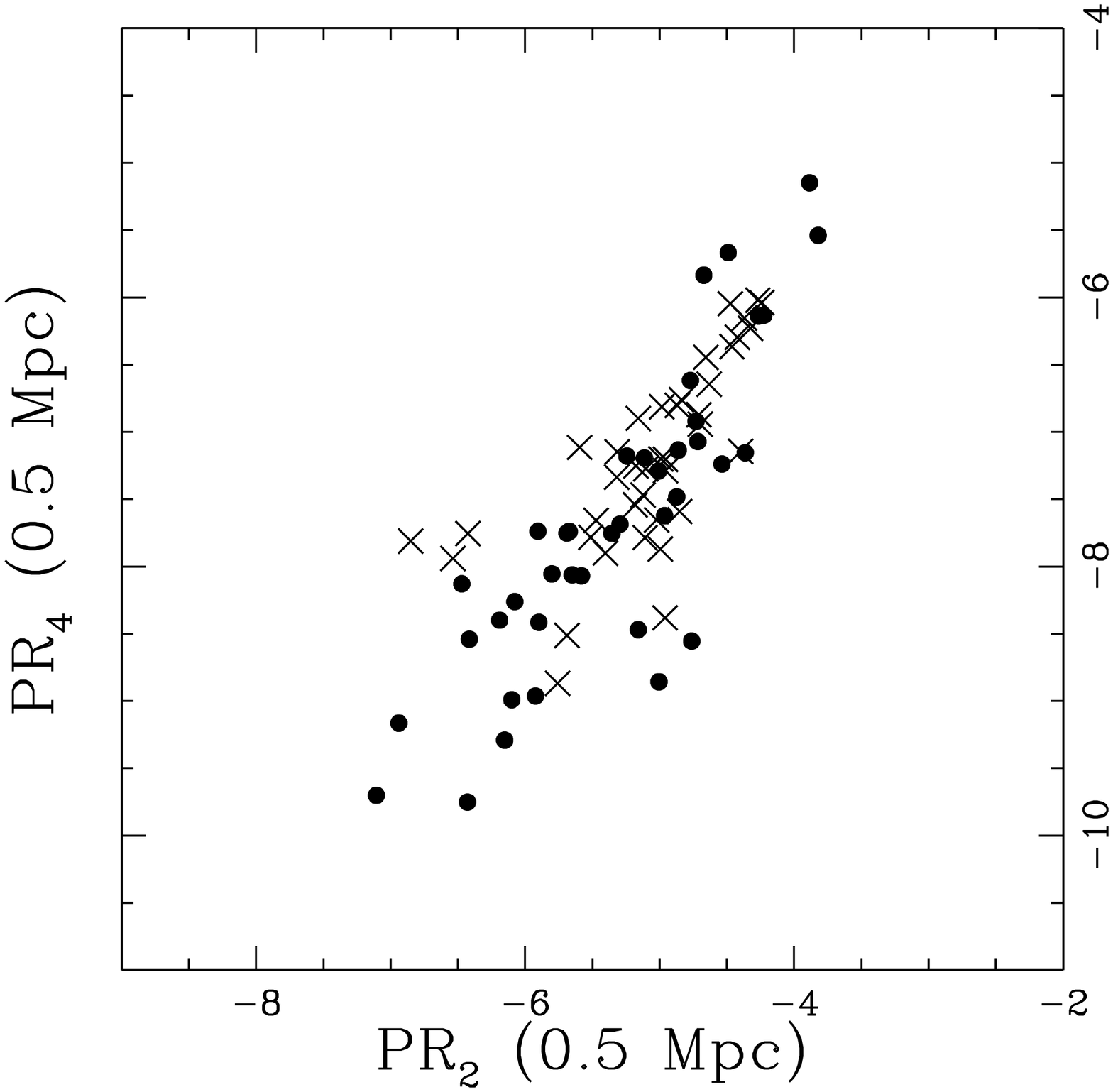}{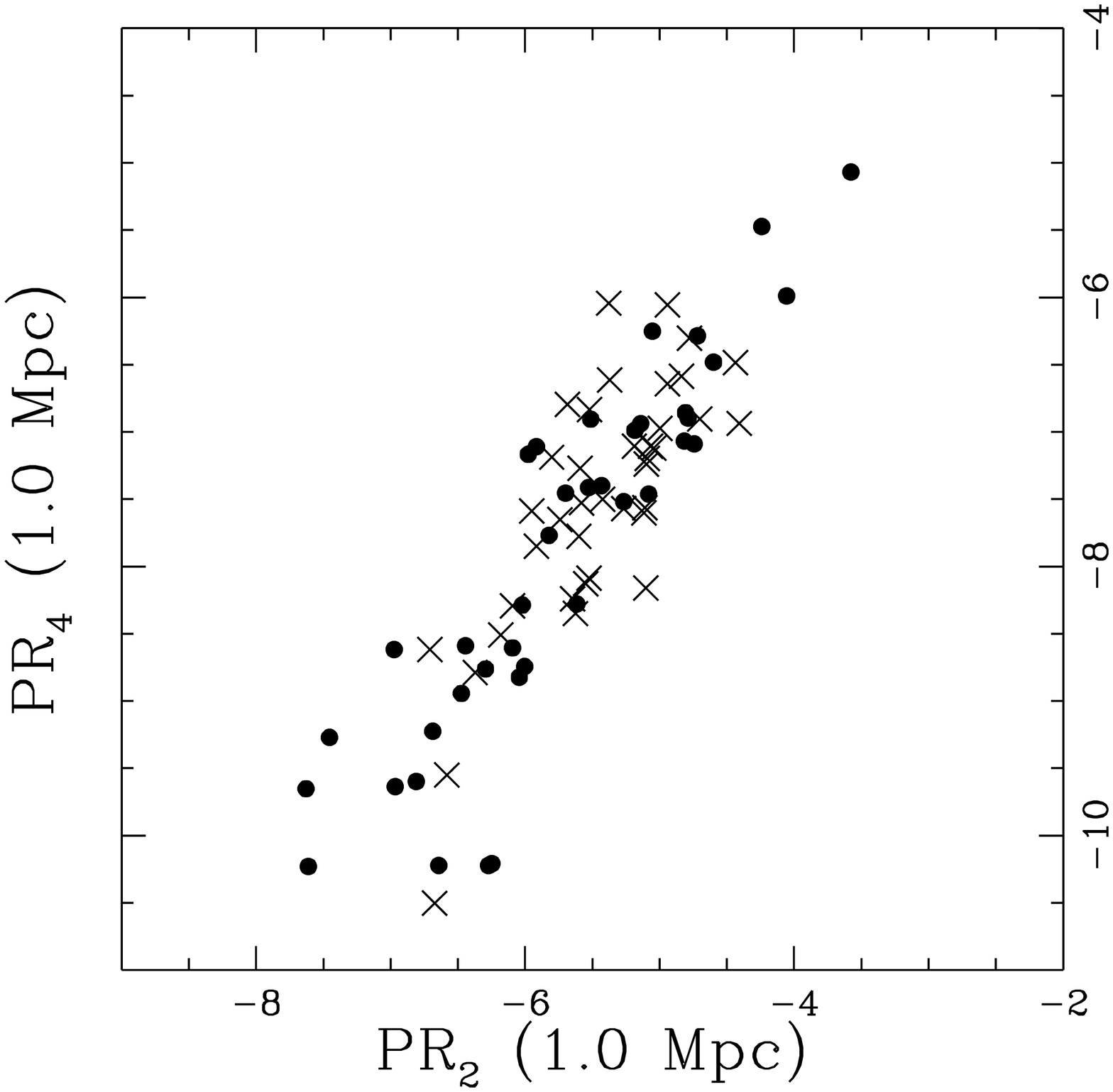}

\end{figure}

\begin{figure}
\caption{  \label{fig.sfhist.05mpc} }

\raggedright 

Histograms for the $PR_m$ in the $0.5h^{-1}_{80}$ Mpc
aperture. Spectral index $n=0$ (SF00) is given by the solid line and
spectral index $n=-2$ (SF20) by the dotted line.

\plottwo{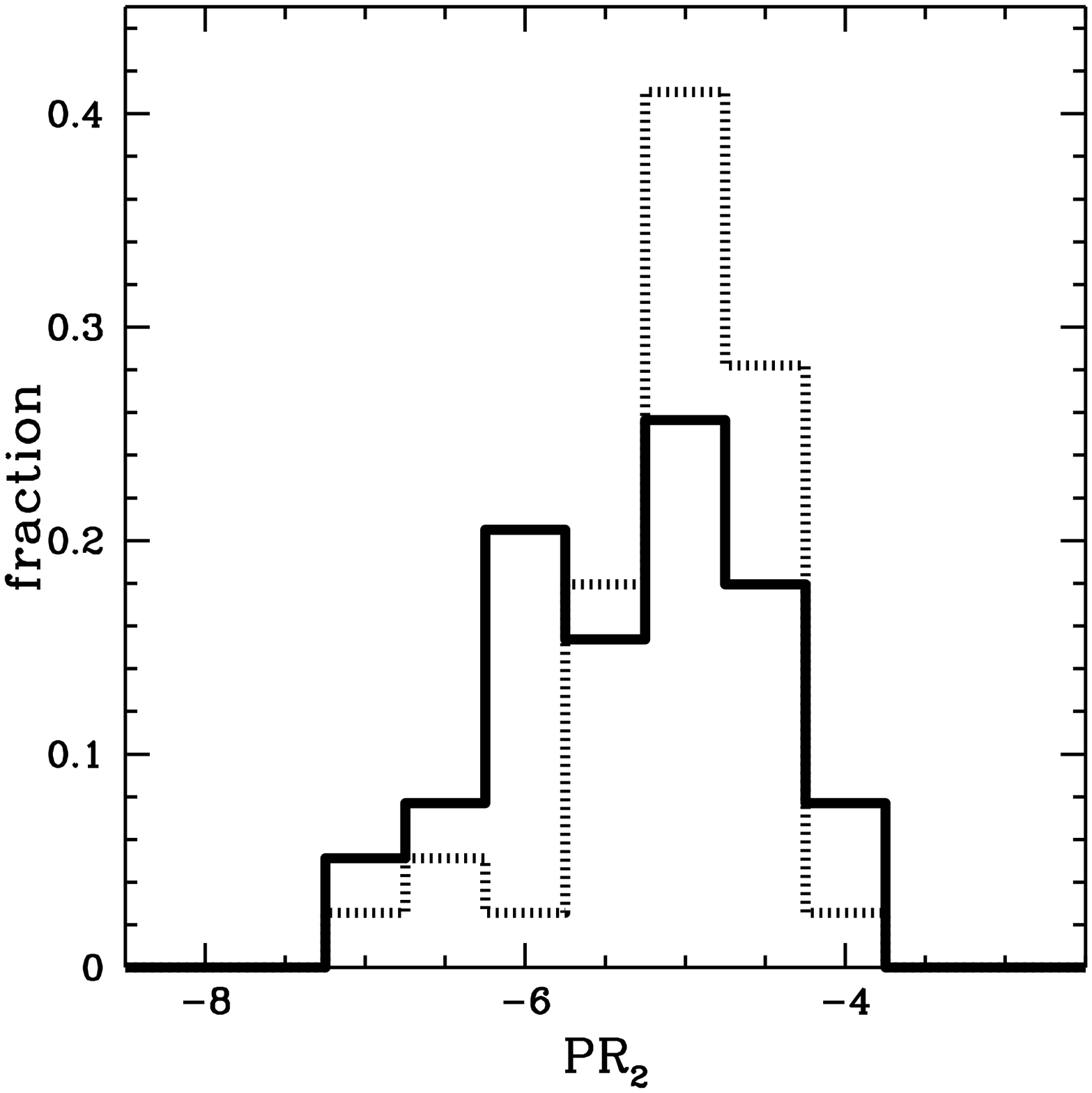}{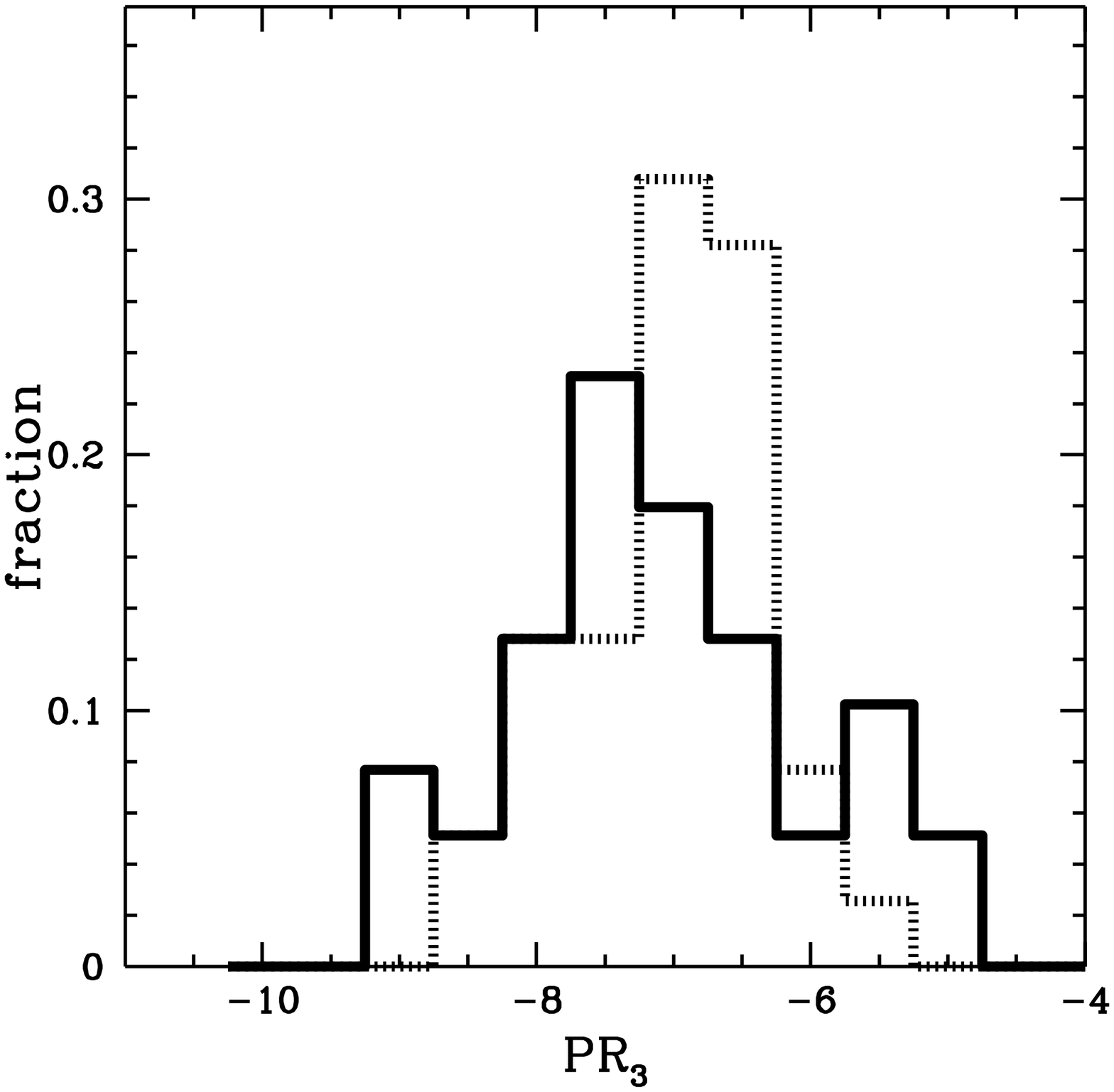}

\end{figure}

\begin{figure}
\caption{  \label{fig.sfhist.10mpc} }

\raggedright As Figure \ref{fig.sfhist.05mpc}, but for the
$1.0h^{-1}_{80}$ Mpc aperture.

\plottwo{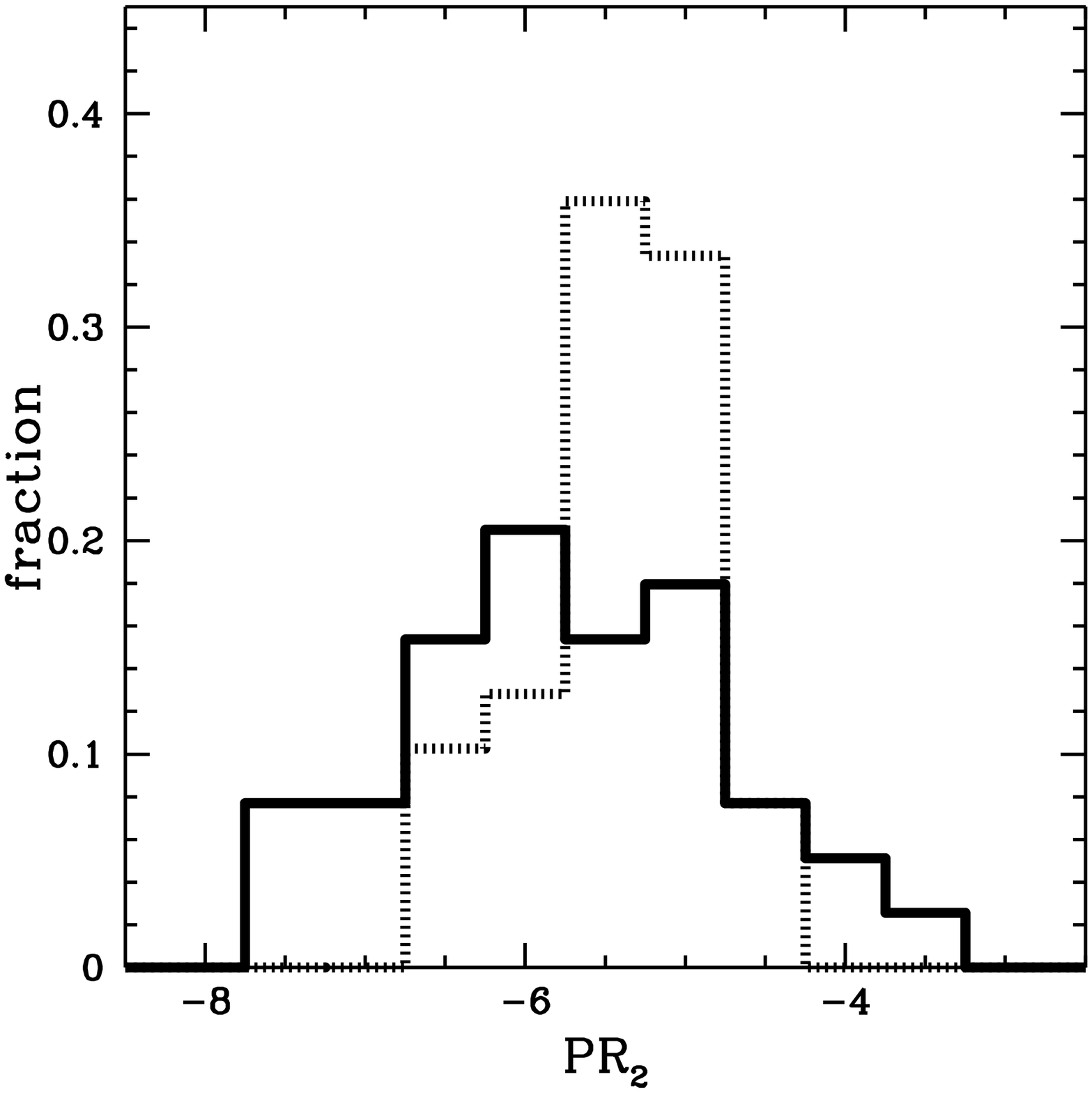}{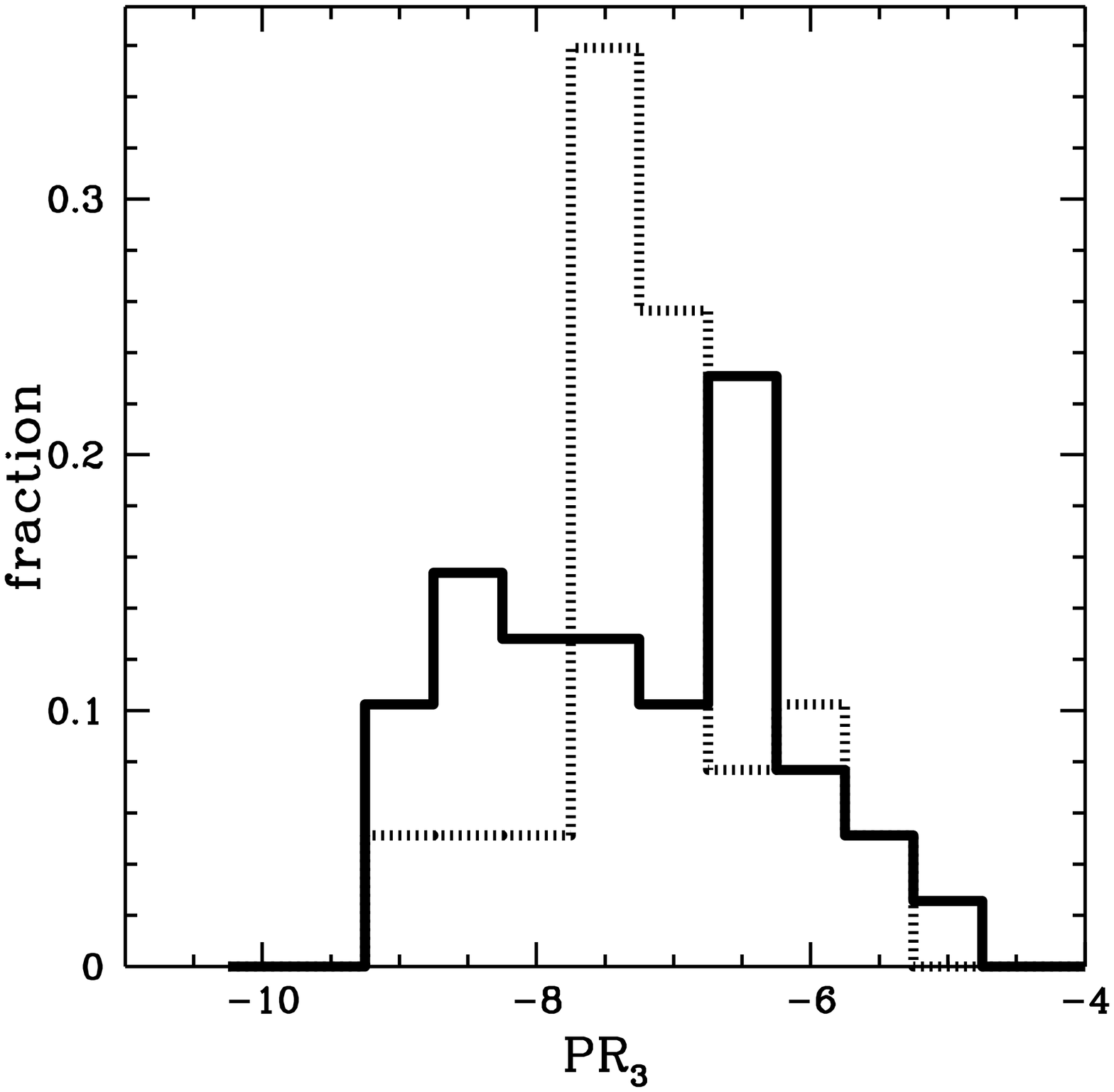}

\end{figure}

\begin{figure}
\caption{  \label{fig.avgpr.pk} }

\raggedright The standard deviation as a function of the average value
of the PRs in the $0.75h^{-1}_{80}$ Mpc aperture for the models in \S
\ref{spectrum}. The error bars represent $1\sigma$ errors estimated from
1000 bootstrap resamplings.

\plottwo{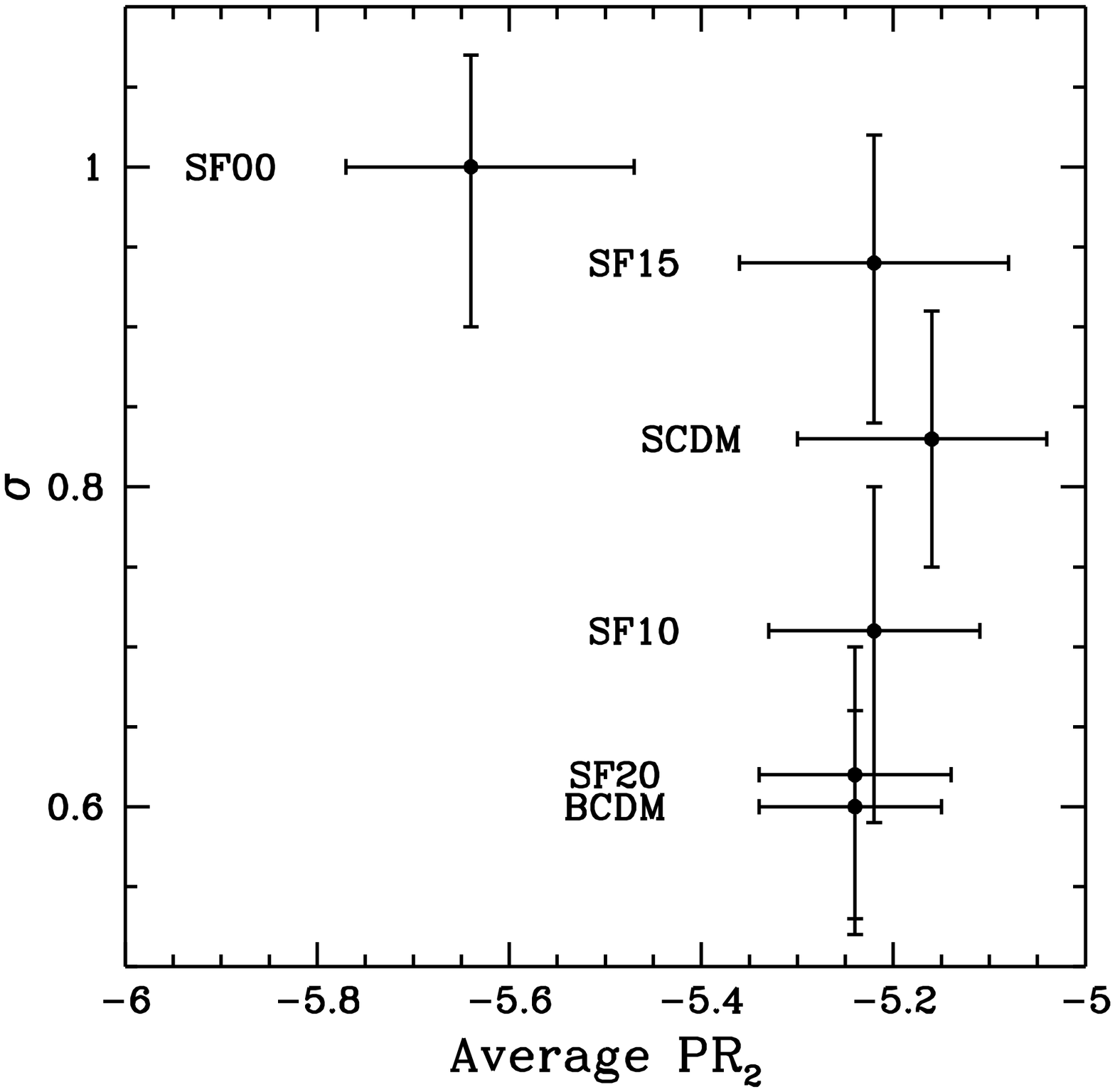}{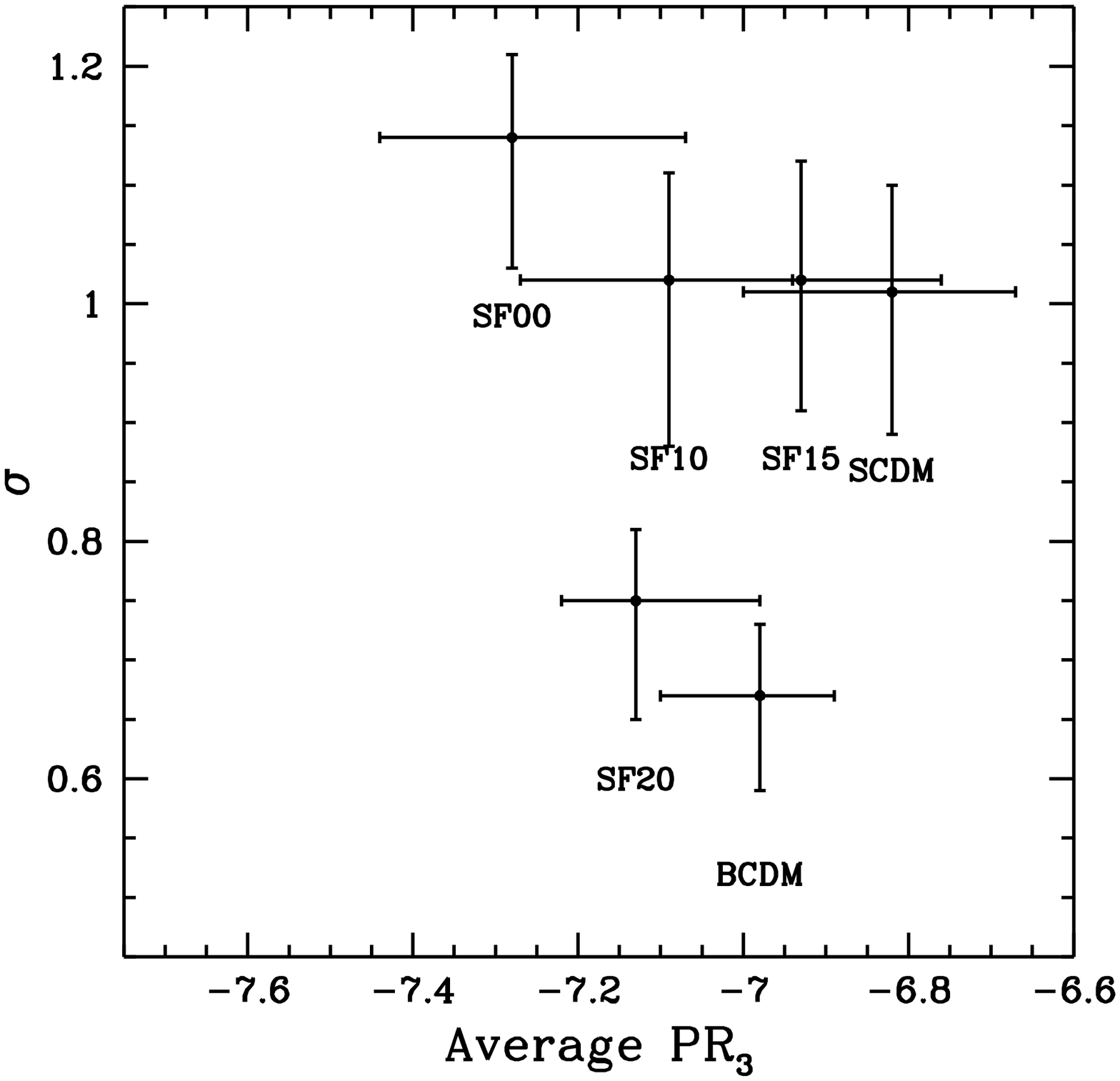}

\end{figure}

In Figure \ref{fig.ets.sf} we plot for the $n=0,-2$ models the PR
correlations for $m=(2,3)$ and $m=(2,4)$ in the $(0.5,1.0)h^{-1}_{80}$
Mpc apertures. Histograms for the individual $PR_m$ in these apertures
are displayed in Figures \ref{fig.sfhist.05mpc} and
\ref{fig.sfhist.10mpc}. Table \ref{table.avgpr} lists the means,
Figure \ref{fig.avgpr.pk} plots the standard deviations versus the
averages of the $PR_m$ in the $0.75h^{-1}_{80}$ Mpc aperture, and
Table \ref{table.test} gives the results of the significance tests.

The means of $PR_3$ are very consistent for the $n=0,-2$ models at all
radii examined. Those for $PR_2$ may show some differences in their
means, with the $n=-2$ models perhaps having systematically smaller
values. The significance of the different means for $PR_2$ is only
formally marginal, with $p_t=(11\%,4\%,10\%)\%$ for aperture radii
$(0.5,0.75,1.0)h^{-1}_{80}$ Mpc.

However, the possible small differences in the means of $PR_2$ are
dwarfed by the corresponding highly significant differences in its
variances. Generally the variances for all the $PR_m$ in all the
apertures are smaller for the $n=-2$ clusters. The most significant
variance differences are observed for $PR_2$ which has $p_F<1\%$ in
$(0.75,1.0)h^{-1}_{80}$ Mpc apertures and $p_F\sim 5\%$ for
$0.5h^{-1}_{80}$ Mpc .  The variances for $PR_3$ show differences
but at a lower level of significance and only in the
$(0.5,0.75,1.0)h^{-1}_{80}$ Mpc apertures; i.e. $p_t=(3\%,1\%,4\%)$.

Similar to what we found in \S \ref{omega}, the differences implied by
the KS test generally follow the significances implied by the t-test
and F-test; i.e. higher order effects in the distributions are
probably not overly important (at least for our sample sizes of 39
clusters). Moreover, again we find that analysis of the $PR_m$ in terms
of the evolutionary tracks does not add useful information to the
previous results. The mean and variance effects for the individual $PR_m$
translate to very similar behavior along $PR_{2-4}$. The direction
perpendicular to $PR_{2-4}$ is essentially consistent for all of the
tests. As a result, we do not emphasize the KS tests or the
evolutionary tracks further.

\subsection{Intermediate $n$}

The behavior for other $n$ is similar, but depends to some extent on
the range examined. We find that the range of $n$ which accentuates
differences in the $PR_m$ is between $n=0,-1$.  Over the range $n=0,-1$
the discrepancy of means for $PR_2$ essentially follows that of the
full $n=0,-2$ discussed in \S \ref{n0n2}.  However, the variances are
not so highly discrepant as before, with $p_F=3\%$ for $PR_2$ for
aperture radii $(0.5,0.75)h^{-1}_{80}$ Mpc; elsewhere the variances of
$PR_2$ are consistent between the $n=0,-1$ models.  Over the
$n=0,-1$ range $PR_3$ is consistent for all statistics at all radii
examined.

The $PR_m$ exhibit very few differences over the range of indices
$n=-1,-2$. For all radii examined the means and KS statistics are
consistent for all the $PR_m$. However, the variances do show some
marginal differences.  The $1.0h^{-1}_{80}$ Mpc apertures has the most
significance difference where $p_F=1.5\%$ for both $PR_2$ and
$PR_3$. Also in the $0.75h^{-1}_{80}$ Mpc aperture $PR_2$ has
$p_F=7\%$. Otherwise the variances of these $PR_m$ are consistent. (We
mention that the variance of $PR_2$ for the $n=-1.5$ model in the
$0.75h^{-1}_{80}$ Mpc aperture lies above that for the $n=-1$ model in
Figure \ref{fig.avgpr.pk}, but the difference is not statistically
significant.)

\subsection{SCDM vs. BCDM\label{svsb}}

Now we consider the $\Omega_0=1$, CDM model with a lower
normalization, $\sigma_8=0.51$, which we refer to as the biased CDM
model, BCDM.  The means and variances for the BCDM model are listed in
Table \ref{table.avgpr}, the standard deviation versus the average
$PR_m$ in the $0.75h^{-1}_{80}$ Mpc aperture are plotted in Figure
\ref{fig.avgpr.pk}, and the results for the significance tests in
comparison to SCDM are given in Table \ref{table.test}.

The means of all the $PR_m$ at all aperture radii are consistent for the
SCDM and BCDM models.  The $PR_m$ variances of the SCDM clusters generally
exceed those of the BCDM clusters. The significance levels of the
differences are only marginal $(p_F\sim 3\%)$ and appear to be most
important in the $0.75h^{-1}_{80}$ Mpc aperture. 

It is possible that the slight variance differences between the SCDM
and BCDM models are due to the difference in resolution between the
two simulations; i.e. the clusters in the BCDM simulations contain
about half the number of particles of the SCDM clusters. We would
expect that the effects of resolution would be most important in the
smallest apertures (which we do observe), although we would probably
expect that the means as well as the variances would be affected
(which we do not observe). We mention that the BCDM model performs
virtually identically to the SCDM model when compared to the OCDM and
LCDM models.

\subsection{Performance Evaluation II.}

The variances of the $PR_m$ show the most significant differences between
models with different power spectra; $PR_2$ generally has the most
sensitive variances over the parameter ranges explored.  Decreasing
$n$ and $\sigma_8$ both decrease the $PR_m$ variances, the differences
being of similar magnitude for the $n=0,-1$ models and the SCDM and
BCDM models.  The means of the $PR_m$ are much less sensitive to the
models with different $n$ and $\sigma_8$, with $PR_2$ showing the
largest significant differences which are always less than differences
in the variances. No significant differences in the means are observed
for $PR_3$ over the range of power spectra studied.

The predominant effect of the power spectrum on the variances of the
$PR_m$ is intriguing. It is reasonable that when the amount of
small-scale structures is reduced (smaller $n$) or the population of
cluster-sized structures is made more uniform (smaller $\sigma_8$)
that the $PR_m$ distributions would also be more uniform.  The observed
low sensitivity of the $PR_m$ means to the power spectra is also reasonable
since on average the $PR_m$ means should only be affected by the rate of
mass accretion through the aperture of radius $R_{ap}$, not by the
sizes of the individual accreting clumps.

\section{Discussion\label{disc}}

In the previous sections we have seen that differences in $\Omega_0$
and $P(k)$ in CDM models are reflected in the spatial morphologies of
clusters when expressed in terms of the $PR_m$.  For the purposes of
probing $\Omega_0$, our analysis indicates that $PR_3$ is the best
PR since its mean is quite sensitive to $\Omega_0$ but very
insensitive to $P(k)$. It is advantageous to also consider $PR_2$
when a cosmological constant is introduced since its means differ for
the SCDM and LCDM models by $\sim 3\sigma$ whereas $PR_3$ only
distinguishes the models at the $\sim 2\sigma$ level.  The marginal
dependence of the mean of $PR_2$ on $P(k)$ is not overly serious
for studying differences in $\Omega_0$ because the differences in
means due to $P(k)$ are always accompanied by larger, more significant
differences in the variances; i.e. different means for $PR_2$ but
consistent variances should reflect differences only in
$\Omega_0$. The best apertures for segregating models are generally
$(0.5,0.75,1.0)h^{-1}_{80}$ Mpc.

A few previous studies have examined the influence of $\Omega_0$ and
$\lambda_0$ on the morphologies of galaxy clusters. Perhaps the most
thorough investigation is that of Jing et. al. (1995) who used N-body
simulations of a variety of CDM models, including versions similar to
our SCDM, OCDM, and LCDM, to study variations of center-shifts and
axial ratios. Jing et. al. reached the same qualitative conclusions as
we do; i.e. the SCDM model is easily distinguished from OCDM and LCDM
because it produces clusters with much more irregular morphologies
than than the others.  However, Jing et al. obtained infinitesimal KS
probabilities for the axial ratio when comparing SCDM to OCDM and
LCDM, a level of significance orders of magnitude different from that
found in this paper. The source of this discrepancy is unclear given
the qualitative similarities of their axial ratio and our $PR_2$.
The disagreement may arise from differences in numerical modeling
between the simulations; i.e.  the results of Jing et al. are derived
from simulations with a larger force resolution ($0.1h^{-1}$ Mpc), and
smaller particle number for the non-SCDM models ($64^3$) than in our
simulations, and have clusters which visually do not show the rich
structures seen in our simulations (Figures \ref{fig.scdm},
\ref{fig.ocdm}, and \ref{fig.lcdm}).

The qualitative results of Jing et al. agree with the hydrodynamic
simulations of Mohr et al. (1995) who also used center shifts and
axial ratios as diagnostics for ``substructure''.  If we visually
estimate the means of the center shifts and axial ratios from Figures
6 and 8 of Jing et al. for their SCDM, OCDM, and LCDM models (actually
OCDM with $\Omega_0=0.2$ and LCDM with $\Omega_0=0.2, \lambda_0=0.8$),
we find that they agree quite well with the corresponding values in
Table 3 of Mohr et al.; i.e. the results from the N-body and
hydrodynamic simulations are very similar, despite the many other
differences between the simulations (e.g., large number of baryons in
OCDM clusters for Mohr et al.).

We can make a similar comparison of the $PR_m$ derived in this paper
with the results from TB who analyzed the small sample of SCDM
clusters formed in the hydrodynamic simulation of Navarro et
al. (1995a). We find that the means (and variances) of the $PR_m$
computed in this paper are very similar to those of the hydrodynamic
clusters; e.g., the mean for $PR_{2-4}$ for $1h^{-1}_{80}$ Mpc may be
read off Figure 7 of TB which shows excellent agreement with the SCDM
value we obtain from the N-body simulations (average
$PR_{2-4}=3.76$). The quantitative similarity between the results,
particularly between the means of the morphological statistics, for
the N-body and hydrodynamic simulations of (Jing et al.,Mohr et al.)
and (this paper,TB) suggest that it is useful to compare the $PR_m$
derived from N-body simulations directly to the X-ray data.

\subsection{Comparison to $ROSAT$ Clusters\label{rosat}}

Among the biases that need to be considered in such a comparison are
the effects of cooling flows (e.g., Fabian 1994), selection, and
noise.  Cooling flows increase the X-ray emission in the cluster
center, which has the effect on the $PR_m$ of essentially decreasing
the core size of the cluster. Judging by the observed core radii of
``regular'' X-ray clusters we would expect at most a factor of $\sim
2$ difference in core radii (e.g., A401 vs. A2029 in Buote \&
Canizares 1996; also see Jones \& Forman 1984).
\footnote{Large cooling flows only appear in clusters with regular
morphologies (e.g., Jones \& Forman 1992; Fabian 1994; BTb).}
Changing the core radius by a factor of 2 typically changes $PR_2$
(for example) by a small fraction of a decade (see Table 6 of BTa);
this behavior, as we show below, is confirmed using a more thorough
treatment.  The issue of biases between X-ray-selected and
mass-selected samples needs to be addressed with hydrodynamical
simulations. The estimated uncertainties of the $PR_m$ for the $ROSAT$
cluster sample of BTb, which take into account noise and unresolved
sources, do not show any clear biases.

\begin{figure}
\caption{  \label{fig.ets.ros} }

\raggedright 

Joint $PR_m$ distributions in the $(0.5,1.0)h^{-1}_{80}$ Mpc apertures
for the $ROSAT$ (crosses) and SCDM (dots) clusters.

\plottwo{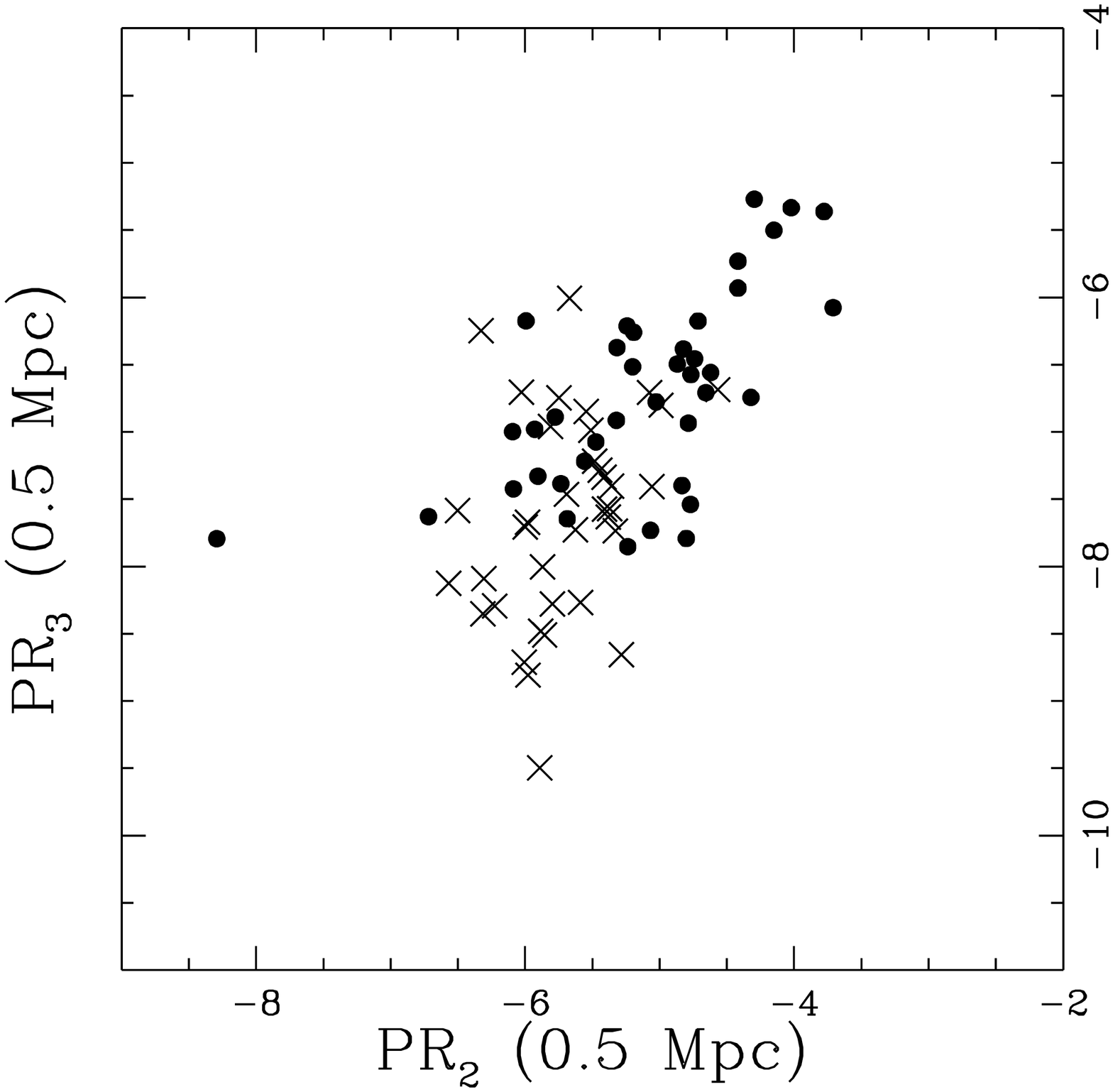}{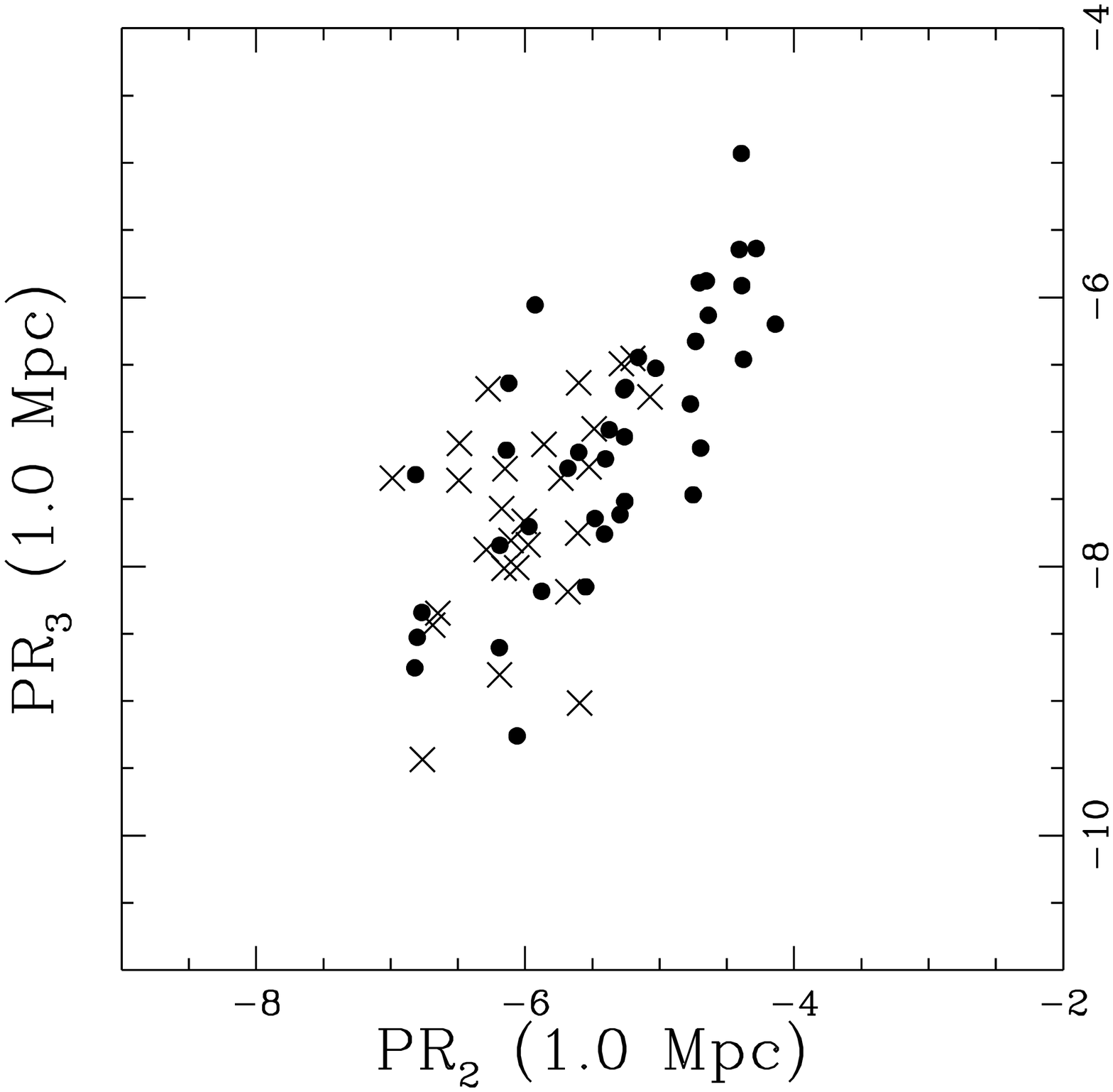}

\plottwo{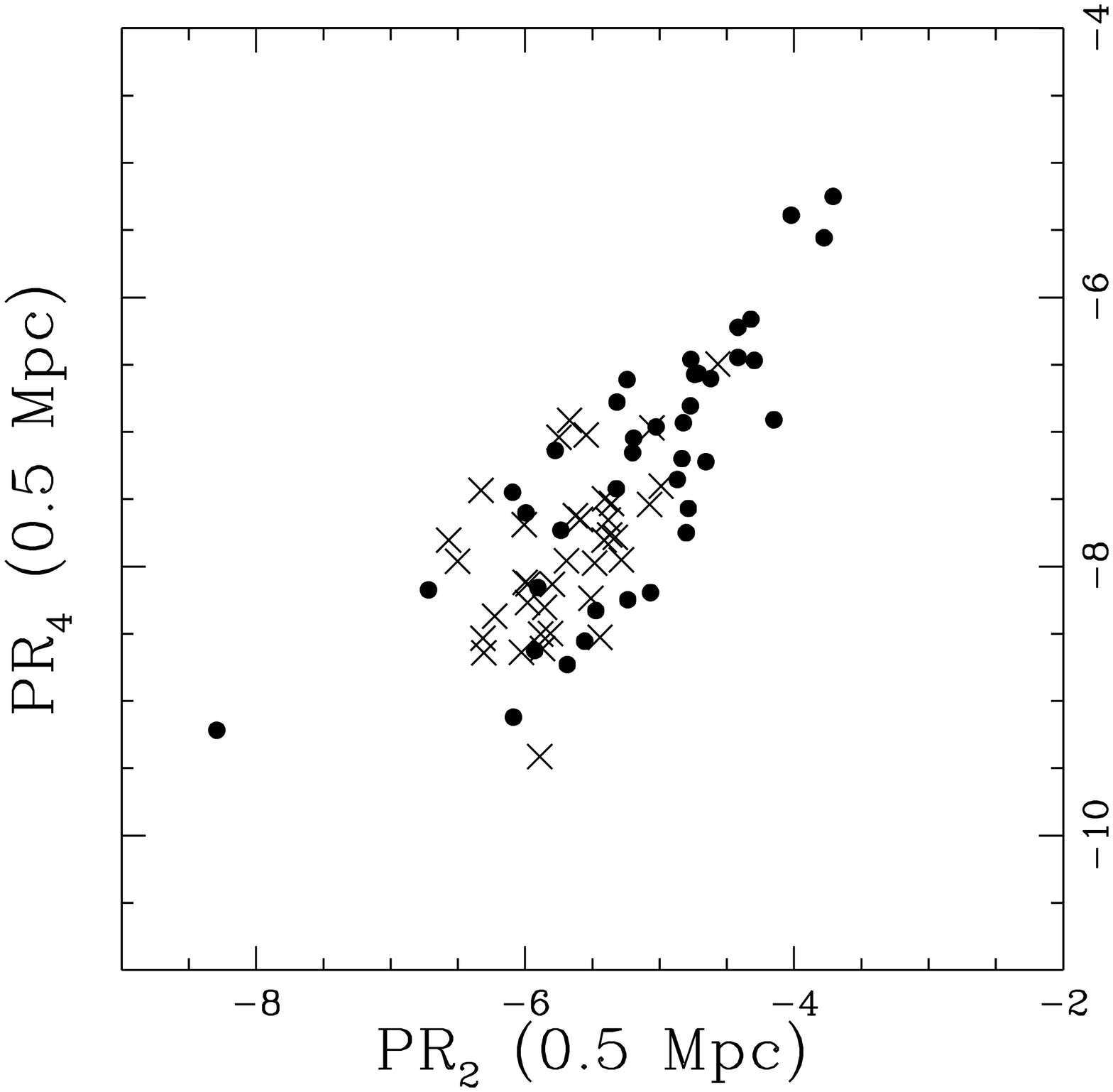}{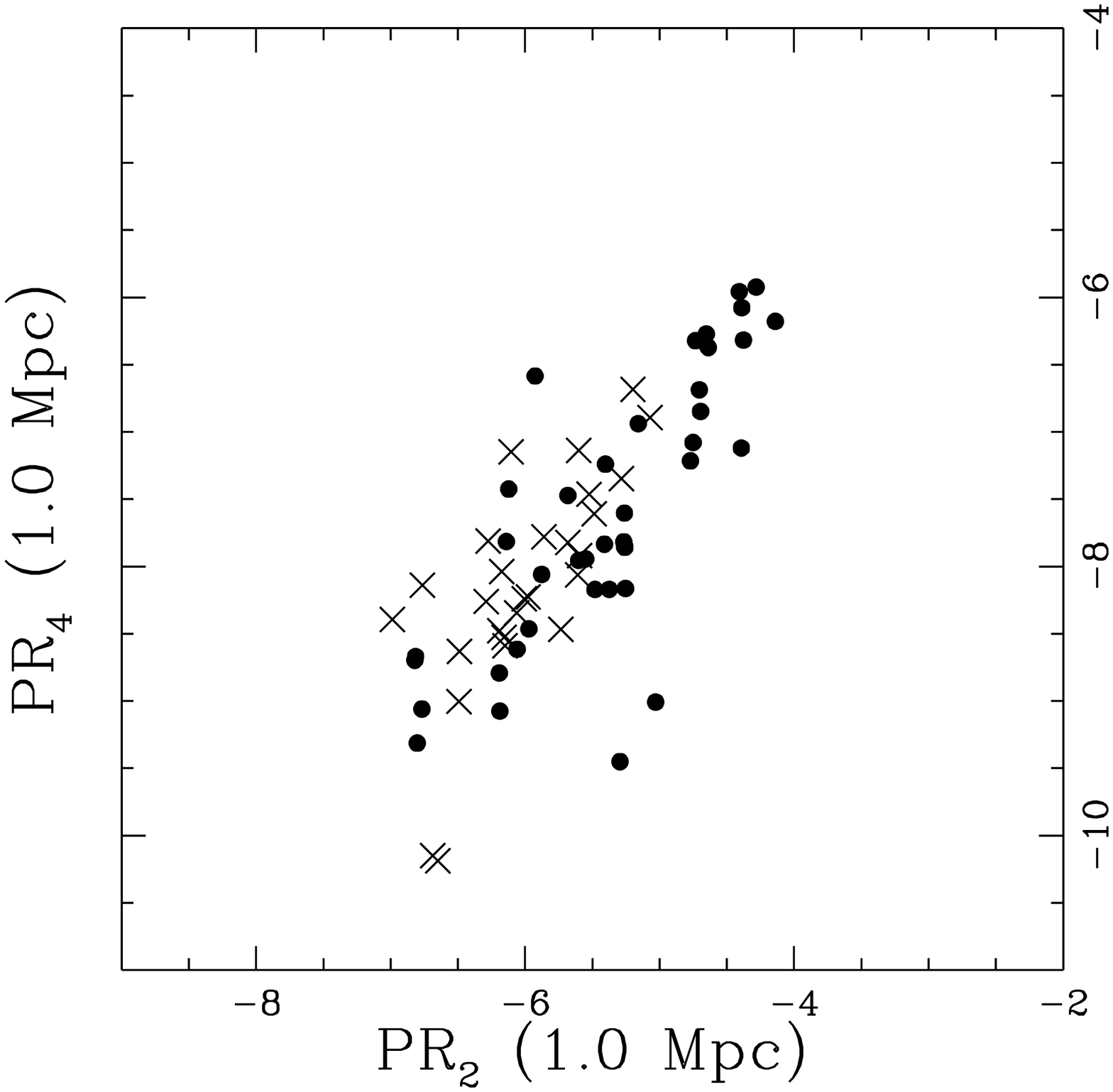}

\end{figure}

\begin{figure}
\caption{  \label{fig.roshist.05mpc} }

\raggedright Histograms for the PRs in the $0.5h^{-1}_{80}$ Mpc
aperture. $ROSAT$ is given by the solid line, SCDM by the dotted line,
and OCDM by the dashed line.

\plottwo{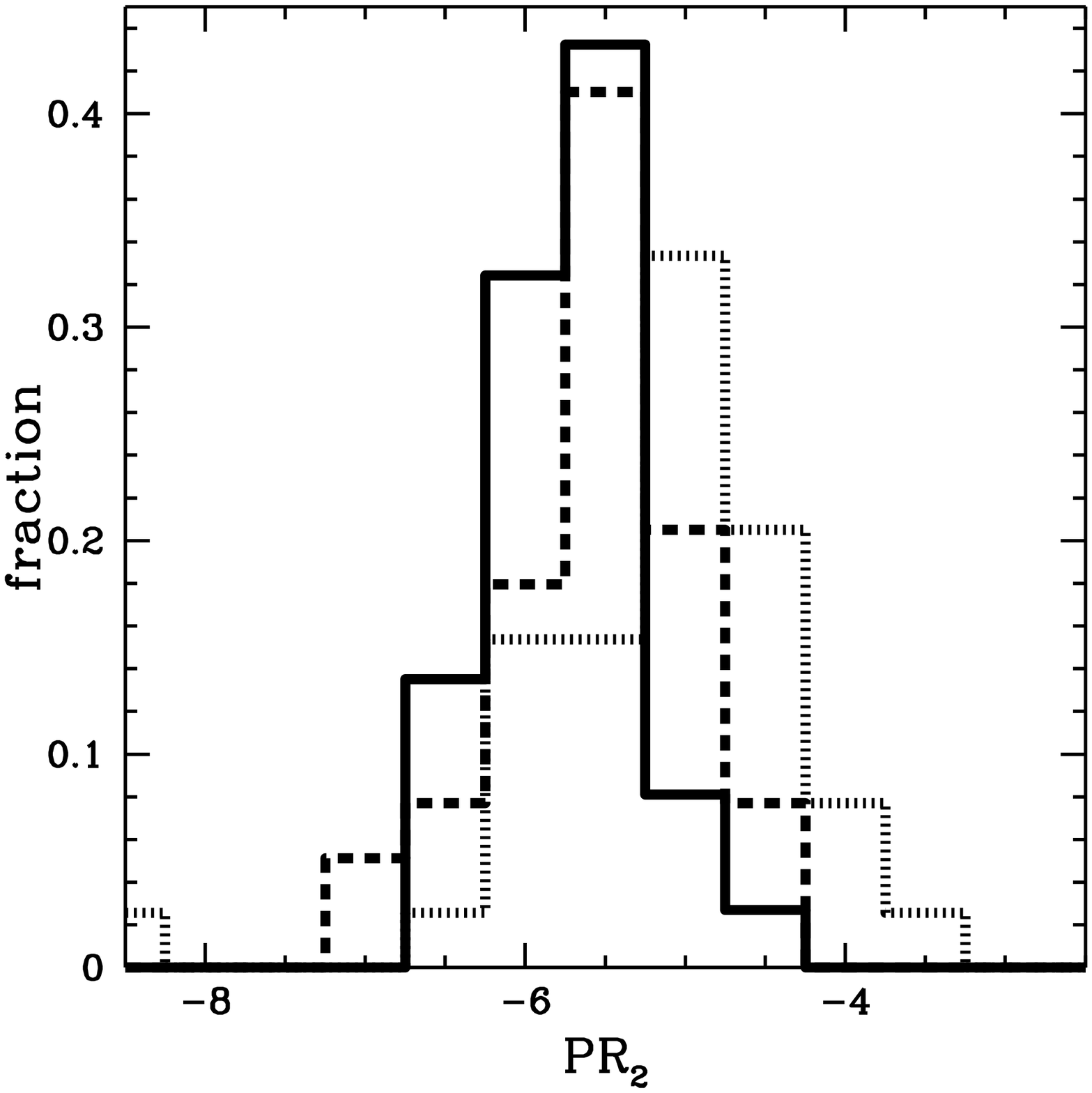}{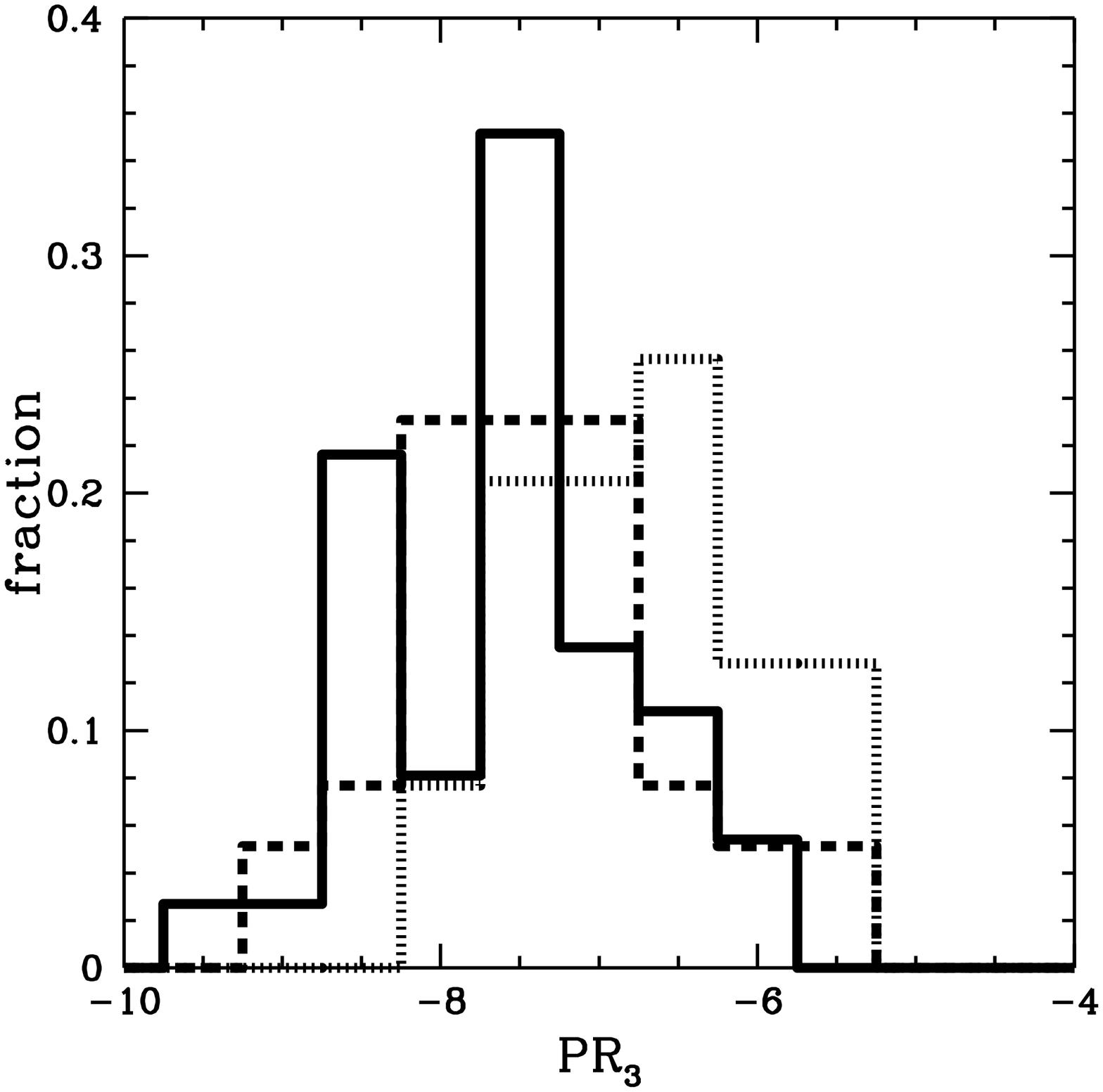}

\end{figure}

\begin{figure}
\caption{  \label{fig.roshist.10mpc} }

\raggedright As Figure \ref{fig.roshist.05mpc}, but for the
$1.0h^{-1}_{80}$ Mpc aperture.

\plottwo{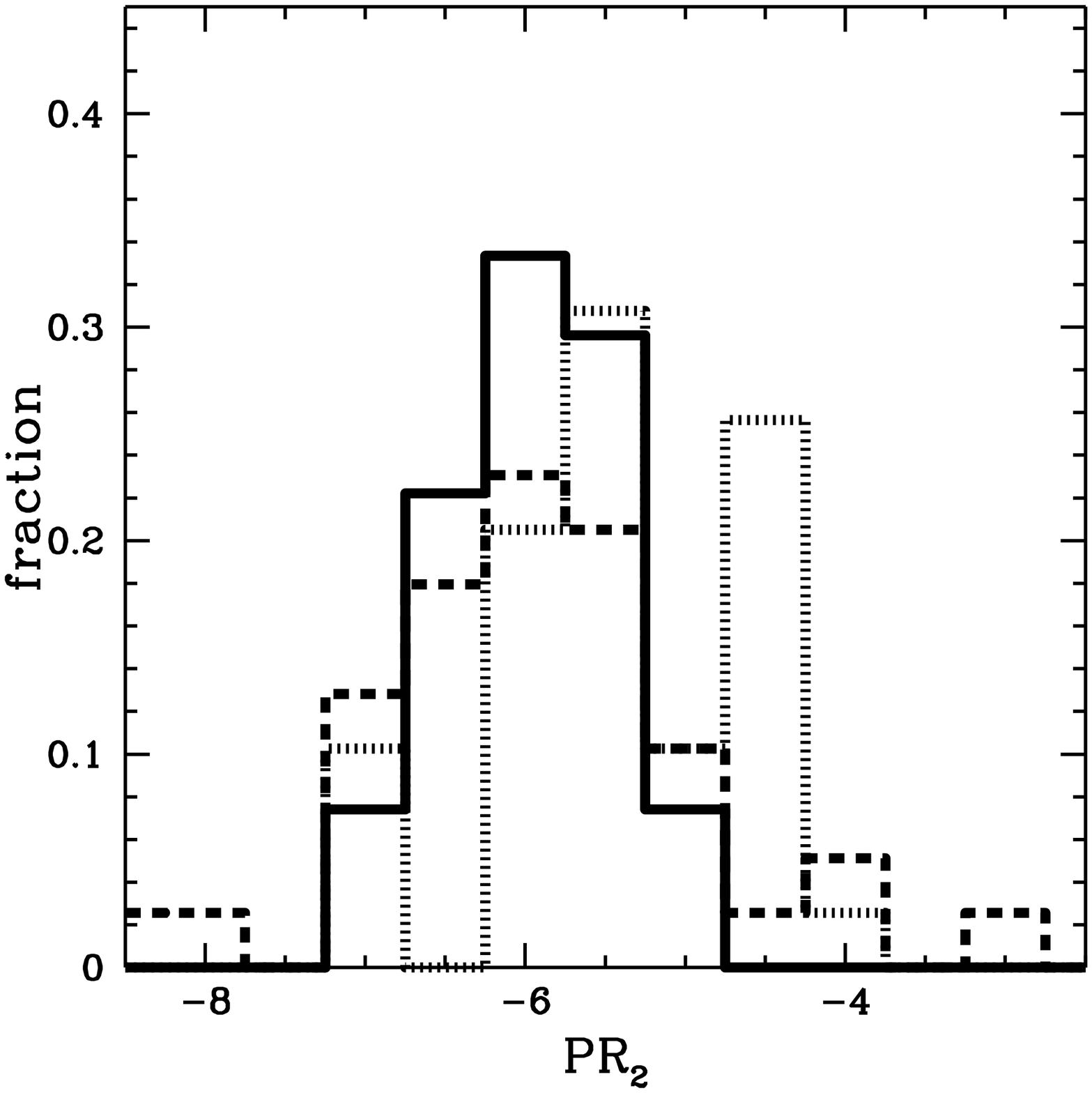}{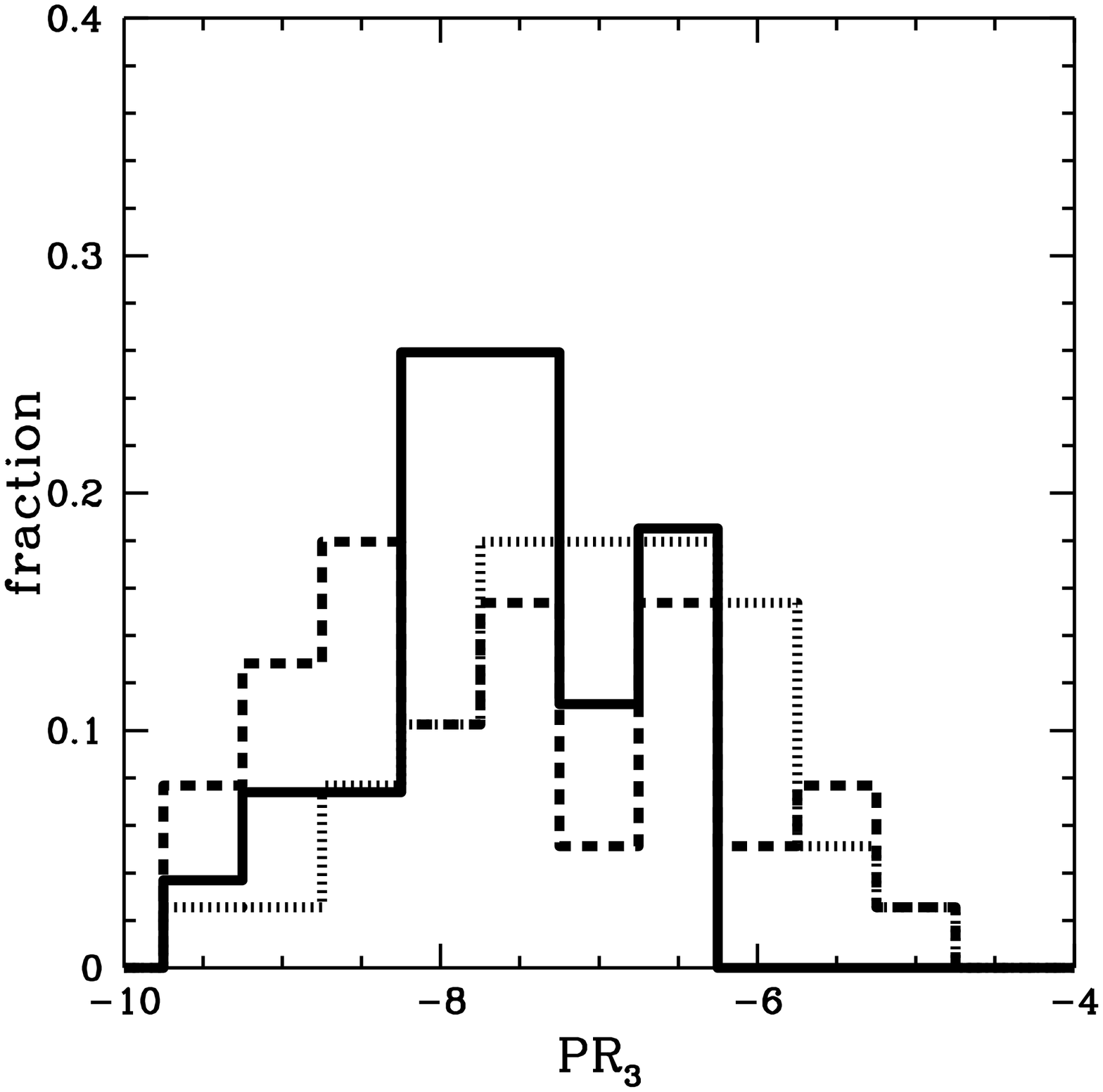}

\end{figure}

\begin{figure}
\caption{  \label{fig.avgpr.rosat} }

\raggedright The standard deviation as a function of the average value
of the PRs in the $0.5h^{-1}_{80}$ Mpc aperture for the ROSAT clusters
and the models discussed in \S \ref{rosat}. The error bars represent
$1\sigma$ errors estimated from 1000 bootstrap resamplings.

\plottwo{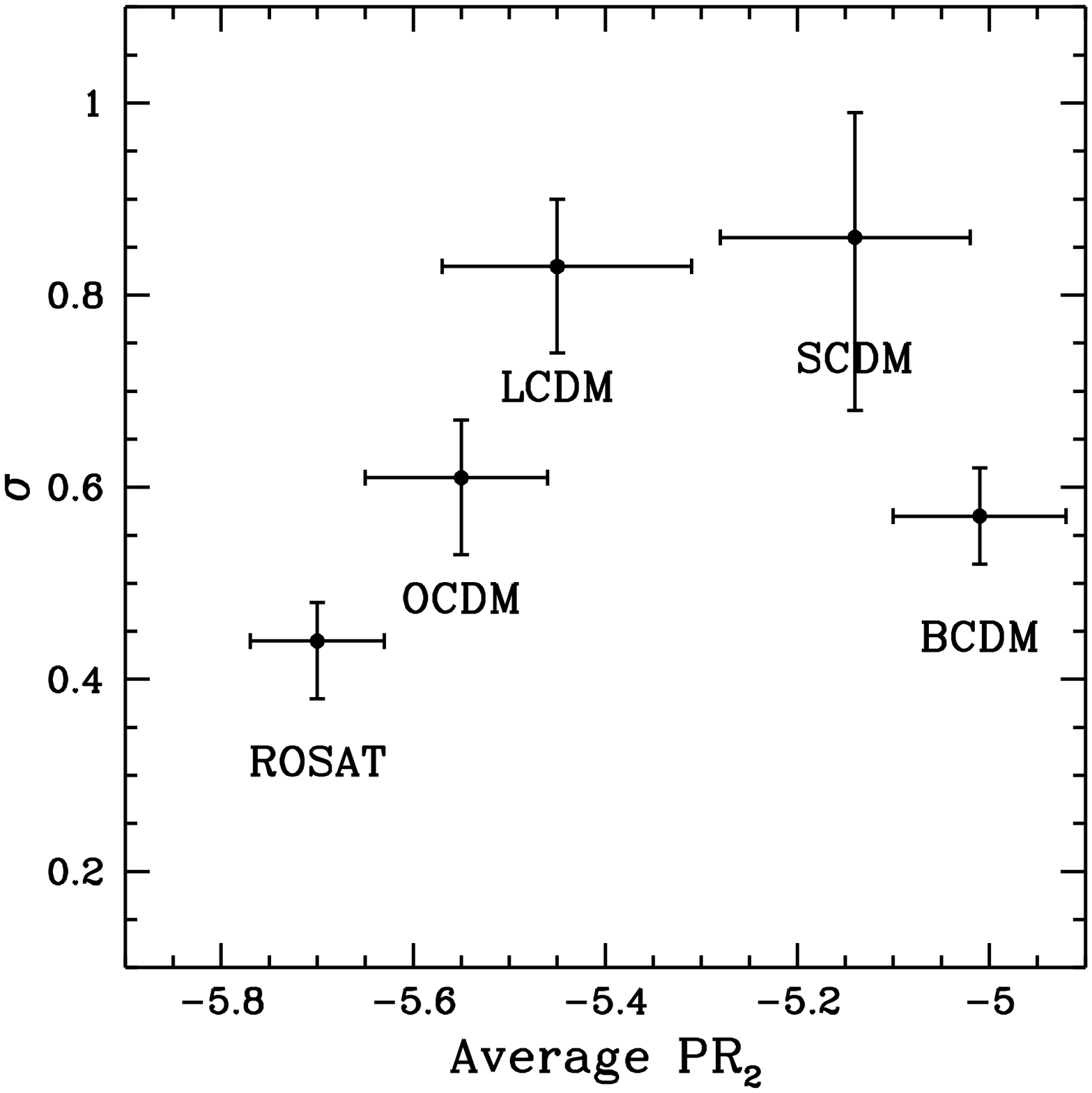}{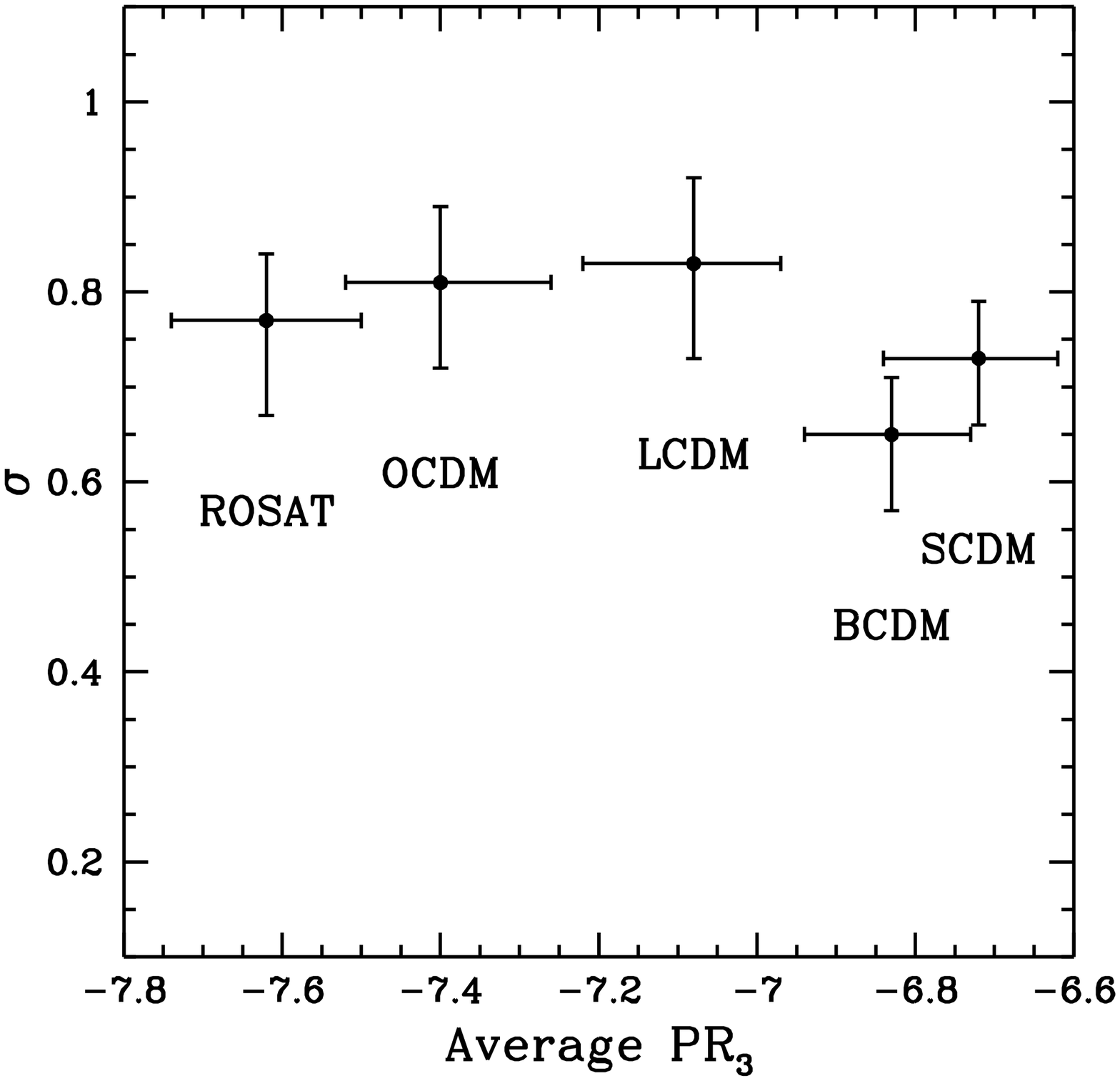}

\end{figure}

In Figure \ref{fig.ets.ros} we display the correlations of the
centroided $PR_m$ for the $ROSAT$ sample of BTb in the
$(0.5,1.0)h^{-1}_{80}$ Mpc apertures; the SCDM clusters are also
plotted for a comparison.  $PR_m$ histograms for these apertures are shown
in Figure \ref{fig.roshist.05mpc} and \ref{fig.roshist.10mpc}, along
with those for the SCDM and OCDM models.  We list the means and
variances for the $ROSAT$ clusters in Table \ref{table.ros.avg}; we
plot in Figure \ref{fig.avgpr.rosat} the standard deviations versus
the means for the $ROSAT$ clusters and models in the $0.5h^{-1}_{80}$
Mpc aperture; the results of the significance tests between the
$ROSAT$ clusters and model clusters are given in Table
\ref{table.ros.test}. We analyze the $ROSAT$ clusters corresponding to
the ``updated Edge et. al. (1990)'' flux-limited sample in BTb which
gives 37 and 27 clusters respectively for the $(0.5,1.0)h^{-1}_{80}$
Mpc apertures; note that all the qualitative features of the results
we obtain below are reproduced when all of the clusters studied in BTb
are used (i.e. 59 and 44 clusters respectively).

The means of the SCDM clusters exceed those of the $ROSAT$ sample to a
high level of significance, with the differences being most pronounced
in the $0.5h^{-1}_{80}$ Mpc aperture. The most significant discrepancy
is for $PR_3$ in the $0.5h^{-1}_{80}$ Mpc aperture for which
$p_t=1.5\times 10^{-4}\%$. The variances for all the $PR_m$ except
$PR_3$ are also significantly different, with the variances of the
SCDM clusters exceeding those of the $ROSAT$ clusters. The SCDM model
has $\sigma_8=1$ which is too high to fit other observations (e.g.,
Ostriker \& Steinhardt 1995).  The BCDM model, which has
$\sigma_8=0.51$, does have $PR_m$ variances in better agreement with the
$ROSAT$ sample. However, the means are in essentially the same level
of disagreement. In fact, $PR_2$ has a much more significant mean
discrepancy $(p_t=1.0\times 10^{-4}\%)$ in the $0.5h^{-1}_{80}$ Mpc
aperture.

In contrast, the $PR_m$ have means that are entirely consistent for the
OCDM and $ROSAT$ clusters in both apertures. The variances of the
centroided $PR_m$ are significantly discrepant, particularly in the
$1.0h^{-1}_{80}$ Mpc aperture, where the OCDM variances exceed the
$ROSAT$ variances. This suggests a lower $\sigma_8$ or $n$ is needed
to bring the variances of the OCDM models into agreement with the
$ROSAT$ sample.  

The $PR_m$ means of the LCDM clusters systematically exceed the $ROSAT$
means, but at a lower level of significance than does SCDM. The
discrepancies are only significant in the $0.5h^{-1}_{80}$ Mpc
aperture, where $PR_3$ $(p_t=0.4\%)$ and $PR^{(pk)}_1$
$(p_t=0.2\%)$ show the most significant discrepancies; the even $PR_m$
show at best a marginal discrepancy in their means $(p_t=10\%-15\%)$.
The variances for the even $PR_m$ are also significantly different,
though only in the $0.5h^{-1}_{80}$ Mpc aperture as well. As the LCDM
and SCDM variances are very similar, we expect that the variance
differences can be largely obviated with a lower value of $\sigma_8$.

The difference in the means of $PR_3$ for the LCDM and $ROSAT$
clusters in the $0.5h^{-1}_{80}$ Mpc aperture, though formally
significant at better than the $3\sigma$ level, represents a shift of
about one-half a decade in $PR_3$; also, when using all 59 clusters
of BTb the significance is only $p_t=4\%$ ($\sim 2\sigma$). As we have
discussed earlier, it is difficult to completely account for a
discrepancy of this magnitude by invoking, e.g., the unsuitability of
the $\rho_{gas}\propto \rho_{DM}$ approximation, observational noise,
or cooling flows.

We may make a more precise estimate of the effects of cooling flows on
the $PR_m$. The ROSAT clusters in the augmented Edge sample all have
estimated mass-flow rates (Fabian 1994) from which we may compute a
luminosity (bolometric) due to the cooling flow following Edge (1989),
$L_{cool} = 3.0\times 10^{41}h^{-2}_{50}\dot{M}T$ erg/s, where
$\dot{M}$ is in $M_{\sun}$/year and $T$ is in keV. Comparing this
cooling luminosity to the total cluster luminosity, $L_{bol}$, using
the results of David et al. (1993) allows us to in effect remove the
cooling gas from the ROSAT $PR_m$.  To a first approximation the cooling
flow affects only $P_0$ because the cooling emission is weighted
heavily towards the aperture center. Hence, to approximately remove
the effects of the cooling flows from the ROSAT clusters we reduce
$P_0$ for each cluster by $(1-L_{cool}/L_{bol})$.  We find that the
$PR_m$ of the ROSAT clusters are modified minimally, the effect being
that the means of the $PR_m$ are increased by $1/10$ of a decade: means
for $PR_2$ and $PR_3$ are -5.60 and -7.52 respectively in the
$0.5h^{-1}_{80}$ aperture; the variances show no significant
systematic effect.  These small mean shifts do reduce the significance
of the LCDM-ROSAT discrepancy, but the discrepancy is still
significant at the $\sim 3\sigma$ level; e.g., $p_t=1.6\%$ for
$PR_3$ and $p_t=1\%$ for $PR^{(pk)}_1$ in the
$0.5h^{-1}_{80}$ aperture, and $p_t=34\%$ for the even $PR_m$.

Although cooling flows alone cannot completely account for the
differences in the ROSAT clusters and the LCDM model, it is very
possible that when combined with the the other effects mentioned above
a sizeable fraction of the half-decade difference could be made up
which would in any event reduce the significance level of the
difference. As a result, we believe the discrepancy of the
LCDM-$ROSAT$ means must be considered preliminary and await
confirmation from appropriate hydrodynamical
simulations.\footnote{This would not necessarily rule out low-density,
flat models having $\Omega_0<0.35$.}

On the other hand, the means of $PR_3$ for the SCDM and BCDM models
exceed the $ROSAT$ means by almost a full decade to a higher formal
significance level $(\sim 4\sigma)$, which in light of the previous
discussion should be considered robust. {\it We conclude that the
$\Omega_0=1$, CDM models cannot produce the observed $PR_m$ of the
$ROSAT$ clusters, and that the discrepancy in $PR_m$ means is due to
$\Omega_0$ being too large.}  This agrees with our conclusions
obtained in TB for the small sample of clusters drawn from the
hydrodynamic simulation of Navarro et al. (1995a). \footnote{If the
small sample of clusters in the Navarro et al. simulation are in fact
biased towards more relaxed configurations at the present day, then
the agreement discussed above between the $PR_m$ computed for the
$\Omega_0=1$ N-body simulations in this paper and the $PR_m$ that TB
computed for the Navarro et al. simulation further strengthens the
SCDM-$ROSAT$ discrepancy.}

Our conclusions are opposite those of Mohr et al. (1995) who instead
concluded that their {\it Einstein} cluster sample favored SCDM over
both OCDM and LCDM. Given the qualitative agreement discussed above
between the Jing et al. and Mohr et al. simulations, as well as
between our present simulations and TB, it would seem that the
discrepancy lies not in the details of the individual
simulations. Moreover, since the centroid shift is qualitatively
similar to our $P^{(pk)}_1/P^{(pk)}_0$, and the axial ratio is
qualitatively related to our $P_2/P_0$, it would seem unlikely that we
would reach entirely opposite conclusions.

The other plausible variable is to consider how BTb and Mohr et
al. computed their statistics on the real cluster data.  The $ROSAT$
data analyzed by BTb have better spatial resolution and sensitivity
than the {\it Einstein} data analyzed by Mohr et al.. This implies
that the Mohr et al. data should be biased in the direction of less
``substructure'' with respect to BTb, which is the opposite of what is
found.  Another important difference between the two investigations is
that the $PR_m$ are computed within apertures of fixed metric size,
whereas Mohr et. al. use a $S/N$ criterion to define the aperture
size. The fixed metric radius used by the $PR_m$ ensures that cluster
structure on the scale $\sim R_{ap}$ is compared consistently which is
not true for the $S/N$ criterion (see BTa); e.g., Mohr et al. use
aperture sizes of $0.38h^{-1}_{80}$ Mpc for Coma and of
$0.81h^{-1}_{80}$ Mpc for A2256.  It is not obvious, however, how this
confusion of cluster scales would explain the discrepancy of our
results with Mohr et al..\footnote{This issue could be addressed by
computing $PR_m$ on the {\it Einstein} sample of Mohr et al., however
such a task is beyond the scope of the present paper.}

\section{Conclusions\label{conc}}

Using the power ratios ($PR_m=\log_{10}(P_m/P_0)$) of Buote \& Tsai
(1995 -- BTa; 1996 -- BTb; Tsai \& Buote 1996 -- TB) we have examined
the sensitivity of galaxy cluster morphologies to $\Omega_0$ and
$P(k)$ using large, high-resolution N-body simulations. X-ray images
are generated from the dark matter by letting the gas density trace
the dark matter. We argue that the $PR_m$ should not be seriously
biased by this approximation because a real gas in hydrostatic
equilibrium with potentials of CDM clusters is rounder, but also has a
larger core radius, the effects of which partially cancel. We also
argue that the approximation should be reasonable during mergers
because of the agreement shown between the evolution of the dark
matter and gas found by TB who analyzed the hydrodynamical simulation
of Navarro et al. (1995a). Finally, The $PR_m$ generated from the
N-body simulations in this paper agree with results from the Navarro
et al. hydrodynamical simulation (TB). Similar agreement is seen
between the results of the N-body simulations of Jing et al. (1995)
and the hydrodynamical simulations of Mohr et al. (1995).

From analysis of several variants of the standard Cold Dark Matter
model, we have shown that the $PR_m$ can distinguish between models
with different $\Omega_0$ and $P(k)$.  Generally, $\Omega_0$
influences the means of the $PR_m$ distributions such that larger
values of $\Omega_0$ primarily imply larger average PR values. The
slope of the power spectrum and $\sigma_8$ primarily influence the
variances of the $PR_m$; smaller $n$ and $\sigma_8$ generally imply
smaller $PR_m$ variances.

For examining $\Omega_0$, our analysis indicates that $PR_3$ is the
best $PR_m$ since its mean is quite sensitive to $\Omega_0$ but very
insensitive to $P(k)$. It is advantageous also to consider $PR_2$
when a cosmological constant is introduced since its means differ for
the SCDM and LCDM models by $\sim 3\sigma$ whereas $PR_3$ only
distinguishes the models at the $\sim 2\sigma$ level.  The dependence
of the mean of $PR_2$ on $P(k)$ is not overly serious for studying
differences in $\Omega_0$ because the differences in means due to
$P(k)$ are always accompanied by larger differences in the variances;
i.e. different means but consistent variances mostly reflect
differences in $\Omega_0$ for $PR_2$. Typically, the best apertures
for segregating models are $(0.5,0.75,1.0)h^{-1}_{80}$ Mpc.

We did not find it advantageous to compare the distributions along and
perpendicular to the ``evolutionary tracks'' in the
$(PR_2,PR_4)$ and $(PR_2,PR_3)$ planes (see BTb and
TB). The distributions along the tracks performed essentially as a
weighted sum of the constituent $PR_m$. The distributions perpendicular
to the tracks were in almost all cases consistent for the
models. Hence, although the evolutionary tracks are useful for
categorizing the dynamical states of clusters, they do not allow more
interesting constraints on $\Omega_0$ and $P(k)$ to be obtained over
the individual $PR_m$. The consistency of the distributions perpendicular
to the evolutionary tracks seems to be a generic feature of the CDM
models.

We compared the $PR_m$ of the CDM models to the $ROSAT$ cluster sample
of Buote \& Tsai (1996). We find that the means of the $\Omega_0=0.35$
OCDM and $ROSAT$ clusters are consistent, but the means of $PR_3$ for
the LCDM and $ROSAT$ clusters are formally inconsistent at the $\sim
3\sigma$ level. We assert that this discrepancy should be considered
marginal due to various issues associated with the simulation --
observation comparison.

However, the means of $PR_3$ for the SCDM and BCDM models (with
$\Omega_0=1$) exceed the $ROSAT$ means by almost a full decade with a
high level of significance $(\sim 4\sigma)$.  Though the formal
significance level of this $\rho^2_{DM}$ / X-ray comparison should be
considered only an approximation, we argue that taking into account
the hydrodynamics and cooling will not reconcile a discrepancy this
large. {\it We conclude that the $\Omega_0=1$ CDM models cannot
produce the observed $PR_m$ of the $ROSAT$ clusters, and that the
discrepancy in $PR_m$ means is due to $\Omega_0$ being too large.}
This agrees with our conclusions obtained in TB for the small sample
of clusters drawn from the hydrodynamic simulation of Navarro et
al. (1995a).  These conclusions are also consistent with other
indicators of a low value of $\Omega_0$ such as the dynamical analyses
of clusters (e.g., Carlberg et al. 1995), the large baryon fractions
in clusters (e.g., White et al. 1993), and the heating of galactic
disks (Toth \& Ostriker 1992).

Our conclusions are inconsistent with those of Mohr et al. (1995) who
instead concluded that their {\it Einstein} cluster sample favored
$\Omega_0=1$, CDM over equivalents of our low-density models, OCDM and
LCDM. We argue that this type of discrepancy is unlikely due to
numerical differences between our simulations. We discuss possible
differences due to how BTb and Mohr et. al. computed their statistics
on the real cluster data.

Large hydrodynamical simulations are necessary to render the
comparison to the $ROSAT$ data more robust. In addition, the effects
of combining data at different redshifts needs to be explored since
cluster formation rates should behave differently as a function of $z$
in different models (e.g., Richstone et. al. 1992). It may also prove
useful to apply $PR_m$ to mass maps of clusters obtained from weak
lensing (Kaiser \& Squires 1993)\footnote{See Wilson, Cole, \& Frenk
(1996), who have recently studied weak-lensing maps obtained from
N-body simulations, and concluded that a ``global quadrupole
statistic'' ($\sim\sqrt{P_2/P_0}$) can distinguish between
low-density and critical density models.}, though for cosmological
purposes it is not clear whether $\rho_{mass}$ will be as responsive
as $\rho^2_{gas}$ to different $\Omega_0$ and $P(k)$.

\acknowledgements

We gratefully acknowledge J. Tsai for his role in facilitating this
collaboration.  We thank E. Bertschinger for suggesting to us the
scheme to normalize the scale-free models using $M_{\star}$, T. Beers
for providing his ROSTAT programs to compute AI and TI, and A. Edge
for providing the expression for $L_{cool}$.  We appreciated the
anonymous referee's prompt reviewing and comments that helped improve
the presentation of the paper.  DAB was supported by grants NASGW-2681
(through subcontract SVSV2-62002 from the Smithsonian Astrophysical
Observatory) and NAG5-2921, and acknowledges the hospitality of the
Institute of Astronomy where the final stages of this work were
carried out.  DAB also expresses gratitude to several senior
scientists who offered encouragement during the early stages of this
project.  GX acknowledges support from NFS HPCC grant ASC93-18185 and
thanks the Pittsburgh Supercomputer Center for use of the CRAY-T3D
machine.

\clearpage
\vfill\eject

\begin{table}[p] 
\caption{Cosmological Models \label{table.models}}
\begin{tabular}{lccccccc} \tableline\tableline \\[-1pt] 
Name & $\Omega_0$ & $\lambda_0$ & $n$ & $\sigma_8$ & $h$ & $z_i$\\
\tableline 
SCDM & 1     &  0     & 1    & 1.00 & 0.5  & 20  \\
OCDM & 0.35  &  0     & 1    & 0.79 & 0.7  & 25     \\
LCDM & 0.35  &  0.65  & 1    & 0.83 & 0.7  & 39      \\
BCDM & 1     &  0     & 1    & 0.51 & 0.5  & 20      \\
SF00 & 1     &  0     & 0    & $\ldots$ & $\ldots$ & $\ldots$   \\
SF10 & 1     &  0     & -1.0 & $\ldots$ & $\ldots$ & $\ldots$  \\
SF15 & 1     &  0     & -1.5 & $\ldots$ & $\ldots$ & $\ldots$  \\
SF20 & 1     &  0     & -2.0 & $\ldots$ & $\ldots$ & $\ldots$  \\ \tableline
\end{tabular}

\tablecomments{$z_i$ is the redshift where the simulations
started. The scale-free models (SF) are normalized to have the same
value of $M_{\star}$ as SCDM (see \S \ref{models}).} 
  
\end{table}

\renewcommand{\arraystretch}{0.75}

{

\begin{table}[p] \small
\caption{Average Power Ratios \label{table.avgpr}}
\begin{tabular}{lrc|rc|rc|rc|rc|rc} \tableline\tableline \\[-1pt] 
& \multicolumn{6}{c}{\large $PR_2$} &
\multicolumn{6}{c}{\large $PR_3$}\\ \\ 
& 
\multicolumn{2}{c}{0.5 Mpc} &
\multicolumn{2}{c}{0.75 Mpc} &
\multicolumn{2}{c}{1.0 Mpc} &
\multicolumn{2}{c}{0.5 Mpc} &
\multicolumn{2}{c}{0.75 Mpc} &
\multicolumn{2}{c}{1.0 Mpc} \\
& avg & $\sigma$ & avg & $\sigma$ & avg & $\sigma$ 
& avg & $\sigma$ & avg & $\sigma$ & avg & $\sigma$ \\ \\[-5pt] \tableline\\[-5pt] 
SCDM &   -5.14  &    0.86 &   -5.16&    0.83&   -5.38  &    0.76  &   -6.72  &    0.73  &   -6.82  &  1.01  &   -7.06  &    0.98\\
OCDM &   -5.55  &    0.61  &   -5.82  &  0.78  & -5.93  & 1.03  &   -7.40  &    0.81  &   -7.56  &    1.13  &   -7.59  &    1.28\\
LCDM &   -5.45  &    0.83  &   -5.69  &    0.85  &   -5.87  &  0.83 &-7.08  & 0.83  &   -7.32  &    1.00  &   -7.38  &    1.02\\
BCDM &   -5.01  &    0.57  &   -5.24  &    0.60  &   -5.41  &  0.63 & -6.83  &    0.65  &   -6.98  &    0.67  &   -7.07  &    0.68  \\
SF00 &   -5.34  &    0.82  &   -5.64  &    1.00  &   -5.75  & 0.97 & -7.12  &    1.02  &   -7.28  &    1.14  &   -7.35  &    1.10  \\
SF10 &   -5.02  &    0.57  &   -5.22  &    0.71  & -5.47  & 0.85 & -6.89  &    0.87  &   -7.09  &    1.02  &   -7.36  &    1.18  \\
SF15 &   -5.20  &    0.85  &   -5.22  &    0.94  & -5.40  & 0.91&   -6.92  &    0.99  &   -6.93  &    1.02  &   -7.05  &    0.96  \\
SF20 &   -5.07  &    0.60  &   -5.24  &    0.62  &   -5.45  & 0.57 & -7.01  & 0.71  & -7.13  & 0.75  &   -7.20  &    0.79  \\ \tableline
\end{tabular}
\tablecomments{Aperture sizes assume h=0.8.}
  
\end{table}

}

\renewcommand{\arraystretch}{1.0}

\renewcommand{\arraystretch}{0.75}

{\scriptsize

\begin{table}[p] \small
\caption{Significance Tests for Power Ratios \label{table.test}} 
\begin{tabular}{ccc|rrr|rrr|rrr} \tableline\tableline \\[-1pt]
&&& \multicolumn{3}{c}{0.5 Mpc} &
\multicolumn{3}{c}{0.75 Mpc} &
\multicolumn{3}{c}{1.0 Mpc}\\[5pt] 
&&& \multicolumn{1}{c}{$p_t$} & \multicolumn{1}{c}{$p_F$} &
\multicolumn{1}{c}{$p_{KS}$} & 
\multicolumn{1}{c}{$p_t$} & \multicolumn{1}{c}{$p_F$} &
\multicolumn{1}{c}{$p_{KS}$} &
\multicolumn{1}{c}{$p_t$} & \multicolumn{1}{c}{$p_F$} &
\multicolumn{1}{c}{$p_{KS}$}\\ 
\multicolumn{3}{c}{Models} & (\%) & (\%) & (\%) & (\%) & (\%) & (\%) & (\%) & (\%) & (\%)\\ \tableline
\tableline \\ \multicolumn{11}{c}{\large $PR_2$}\\ \\
\tableline \\[-5pt] 
SCDM &vs. &OCDM  &   1.63 &  4.09 &   0.94  &  0.06 & 71.71 &  0.07 &  0.89 &  7.18 &  0.94 \\
SCDM &vs. &LCDM  &  10.47 & 87.90 &  21.79  &  0.74 & 89.05 &  1.97 &  0.83 & 58.78 &  3.90 \\
OCDM &vs. &LCDM  &  54.64 &  5.79 &  51.42  & 47.52 & 61.72 & 21.79 & 78.26 & 20.53 & 34.56 \\
SCDM &vs. &BCDM  &  43.06 &  1.51 &  98.09  & 65.19 &  4.65 & 21.79 & 83.52 & 24.87 & 34.56 \\
SF00 &vs. &SF20  &  10.76 &  6.35 &   7.31  &  4.19 &  0.38 &  0.94 & 10.44 &  0.13 &  3.90 \\
\tableline \\ \multicolumn{11}{c}{\large $PR_3$}\\ \\
\tableline \\[-5pt]   
SCDM &vs. &OCDM  &   0.02 & 52.99 &   0.18  &  0.31 & 46.71 &  0.94 &  4.43 & 10.64 &  7.31 \\
SCDM &vs. &LCDM  &   4.44 & 43.91 &  21.79  &  3.26 & 97.10 &  7.31 & 15.88 & 81.58 & 51.42 \\
OCDM &vs. &LCDM  &   9.10 & 88.39 &  21.79  & 30.99 & 44.52 & 51.42 & 43.51 & 16.61 & 51.42 \\
SCDM &vs. &BCDM  &  49.58 & 47.26 &  70.85  & 40.62 &  1.39 & 34.56 & 97.46 &  2.48 & 70.85 \\
SF00 &vs. &SF20  &  57.65 &  2.89 &  21.79  & 49.89 &  1.28 & 21.79 & 50.92 &  4.31 & 12.97 \\
\tableline
\end{tabular}

\tablecomments{Aperture sizes assume h=0.8.}
  
\end{table}

}

\renewcommand{\arraystretch}{1.0}

\begin{table}[p]
\caption{PR Statistics for $ROSAT$ Clusters \label{table.ros.avg}}
\begin{tabular}{lrr|rr} \tableline\tableline \\[-1pt]
& \multicolumn{2}{c}{0.5 Mpc} &
\multicolumn{2}{c}{1.0 Mpc} \\[5pt] 
& avg & $\sigma$ & avg & $\sigma$\\ \tableline
$PR_2$ & -5.70&  0.44& -6.00&  0.50\\  
$PR_3$ & -7.62&  0.77& -7.61&  0.77\\  
\\[-10pt] \tableline
\end{tabular}

\tablecomments{Aperture sizes assume h=0.8.}
  
\end{table}

{\small

\begin{table}[p] \small
\caption{Significance Tests for $ROSAT$ Clusters \label{table.ros.test}}
\begin{tabular}{llll|lll} \tableline\tableline \\[-1pt]
& \multicolumn{3}{c}{0.5 Mpc} &
\multicolumn{3}{c}{1.0 Mpc} \\[5pt] 
Models & $p_t(\%)$ & $p_F(\%)$ & $p_{KS}(\%)$ & $p_t(\%)$ & $p_F(\%)$ &
$p_{KS}(\%)$\\ \tableline
\tableline \\ \multicolumn{7}{c}{\large $PR_2$}\\ \\ \tableline \\[-5pt]
SCDM & 0.60E-01&  0.12E-01&  0.68E-03&	 0.14E-01&  0.24E+01&  0.33E-01\\
BCDM & 0.10E-04&  0.12E+02&  0.56E-04&	 0.12E-01&  0.20E+02&  0.12E-01\\
OCDM & 0.23E+02&  0.52E+01&  0.16E+02&	 0.68E+02&  0.23E-01&  0.43E+02\\
LCDM & 0.11E+02&  0.21E-01&  0.13E+01&	 0.41E+02&  0.69E+00&  0.23E+02\\
\tableline \\ \multicolumn{7}{c}{\large $PR_3$}\\ \\ \tableline \\[-5pt]	   	      		 		  	     
SCDM& 0.15E-03&  0.71E+02&  0.69E-01&	 0.17E+01&  0.20E+02&  0.10E+02\\
BCDM& 0.64E-03&  0.28E+02&  0.32E-02&	 0.35E+00&  0.46E+02&  0.12E+02\\
OCDM& 0.22E+02&  0.80E+02&  0.42E+02&	 0.93E+02&  0.83E+00&  0.29E+02\\
LCDM& 0.44E+00&  0.69E+02&  0.97E+00&	 0.33E+02&  0.14E+02&  0.32E+02\\
\tableline
\end{tabular}

\tablecomments{Aperture sizes assume h=0.8.}
  
\end{table}

}


\clearpage

\topmargin -50 pt

\topmargin 0 pt

\clearpage
\begin{references}

\reference{A} Bertschinger, E., \& Gelb, J. M. 1991, Computers in
Physics, 5, 164 

\reference{A} Bird, C. M., \& Beers, T. C. 1993, \aj, 105, 1596

\reference{A} Buote, D. A., \& Canizares, C. R. 1996, \apj, 457, 565

\reference{A} Buote, D. A., \& Tsai, J. C. 1995a, \apj, 439, 29

\reference{A} Buote, D. A., \& Tsai, J. C. 1995b, \apj, 452, 522 (BTa)

\reference{A} Buote, D. A., \& Tsai, J. C. 1996, \apj, 458, 27 (BTb)

\reference{A} Carlberg, R., Yee, H. K. C., Ellingson, E., Abraham, R.,
Gravel, P., Morris, S., \& Pritchet, C. J. 1995, (astro-ph/9509034)

\reference{A} Lacey, C., \& Cole, S. 1996, \mnras, in press
(astro-ph/9510147) 

\reference{A} Coles, P., \& Ellis, G. 1994, Nature, 370, 609

\reference{A} David, L. P., Slyz, A., Jones, C., Forman, W., \& Vrtlek,
S. D. 1993, \apj, 412, 479

\reference{A} Dekel, A. 1994, \araa, 32, 371

\reference{A} Edge, A. C. 1989, Ph.D. thesis, University of Leicester

\reference{A} Edge, A. C., Stewart, G. C., Fabian, A. C., \& Arnaud,
K. A. 1990, \mnras, 245, 559

\reference{A} Evrard, A. E., Mohr, J. J., Fabricant, D. G., \& Geller,
M. J. 1993, \apjl, 419, 9 

\reference{A} Fabian, A. C. 1994, \araa, 32, 277

\reference{A} Frenk, C. S., Evrard, A., Summers, F., \& White,
S. D. M. 1996, \apj, in press.

\reference{A} Hernquist, L., \& Katz, N. 1989, \apjs, 70, 419

\reference{A} Jing, Y. P., Mo, H. J., B{$\ddot{\rm o}$}rner, G., \& Fang,
L. Z. 1995, \mnras, in press (astro-ph/9412072)

\reference{A} Jones, C., \& Forman W. 1992, in Clusters and Superclusters
of Galaxies (NATO ASI Vol. 366), ed. A. C. Fabian,
(Dordrecht/Boston/London: Kluwer), 49 

\reference{A} Kaiser, N., \& Squires, G. 1993, \apj, 404, 441

\reference{A} Katz, N., \& White, S. D. M. 1993, \apj, 412, 455 

\reference{A} Kauffmann, G., \& White, S. D. M. 1993, \mnras, 261, 921

\reference{A} Mohr, J. J., Evrard, A. E., Fabricant, D. G., \& Geller,
M. J. 1995, \apj, in press

\reference{A} Nakamura, F. E., Hattori, M., \& Mineshige, S. 1995, \aap, in
press

\reference{A} Navarro, J. F., Frenk, C. S., \& White, S. D. M. 1995a, \mnras,
275, 720 

\reference{A} Navarro, J. F., Frenk, C. S., \& White, S. D. M. 1995b,
\apj, submitted (astro-ph/9508025)

\reference{A} Ostriker, J. P., 1993, \araa, 31, 689

\reference{A} Ostriker, J. P., \& Steinhardt, P. J. 1995, Nature, 377,
600

\reference{A} Padmanabhan, T. 1993, Structure Formation in the
Universe (Cambridge: Cambridge Univ. Press)

\reference{A} Press, W. H., Teukolsky, S. A., Vetterling, W. T., \&
Flannery, B. P. 1995, Numerical Recipes (Cambridge: Cambridge
Univ. Press) 

\reference{A} Richstone, D. O., Loeb, A., \& Turner, E. L. 1992, \apj,
393, 477

\reference{A} Tsai, J. C., \& Buote, D. A. 1996, \mnras, in press
(astro-ph/9510057) (TB)

\reference{A} Toth, G., \& Ostriker, J. P. 1992, \apj, 389, 5

\reference{A} White, S. D. M., Navarro, J. F., Evrard, A. E., \& Frenk, C. S.
1993, Nature, 366, 429

\reference{A} Wilson, G., Cole, S., \& Frenk, C. S. 1996, \mnras,
submitted (astro-ph/9601110)

\reference{A} Xu., G. 1995a, Ph.D. thesis, Princeton University

\reference{A} Xu., G. 1995b, \apjs, 98, 355

\end{references}
\end{document}